\documentclass{article}
\usepackage[margin=1.25in]{geometry}
\usepackage{graphicx}
\usepackage{amsmath}
\usepackage{amssymb}
\usepackage{mathtools}
\usepackage{parskip}
\usepackage[hidelinks]{hyperref}
\usepackage{graphicx}
\usepackage[bf]{caption}
\usepackage{subcaption}

\usepackage{color}
\usepackage{amssymb}
\usepackage{natbib}
\newtheorem{remark}{Remark}[section]
\begin{document}
\title{
	On modelling bicycle power for velodromes: Part~II\\
	{\Large Formulation for individual pursuits}
}
\author{
	Len Bos%
	\footnote{
		Universit\`a di Verona, Italy, \texttt{leonardpeter.bos@univr.it}
	}\,,
	Michael A. Slawinski%
	\footnote{
		Memorial University of Newfoundland, Canada, \texttt{mslawins@mac.com}
	}\,,
	Rapha\"el A. Slawinski%
	\footnote{
		Mount Royal University, Canada, \texttt{rslawinski@mtroyal.ca}
	}\,,
	Theodore Stanoev%
	\footnote{
		Memorial University of Newfoundland, Canada, \texttt{theodore.stanoev@gmail.com}
	}
}
\date{January~8, 2021}
\maketitle
\begin{abstract}
We model the instantaneous power on a velodrome\,---\,as applied to individual pursuits and other individual time trials\,---\,taking into account its straights, circular arcs, and connecting transition curves.
The forces opposing the motion are air resistance, rolling resistance, lateral friction and drivetrain resistance.
We examine the constant-cadence and constant-power cases, and discuss their results, including an examination of an empirical adequacy of the model.
We also examine changes in the kinetic and potential energy.
\end{abstract}
\section{Introduction}
This article is a continuation of research presented by \citet{DSSbici1}, \citet{DSSbici2}, \citet{BSSbici6} and, in particular, by \citet{SSSbici3}.
Herein, using a mathematical model, we examine the power required on a velodrome, for an individual pursuit.
It can be also applied to other individual races, such the kilometer time trial and the hour record; the essential quality is the constancy of effort.
In each case, the opposing forces consist of air resistance, rolling resistance, lateral friction and drivetrain resistance.
We consider a velodrome with its straights, circular arcs, and connecting transition curves, whose inclusion\,---\,while presenting a certain challenge, and neglected in previous studies \citep[e.g.,][]{SSSbici3}\,---\,increases the empirical adequacy of the model.
Herein, a model is empirically adequate if it accounts for measurements~\citep{Fraassen}.

We begin this article by expressing mathematically the geometry of both the black line%
\footnote{The circumference along the inner edge of this five-centimetre-wide line\,---\,also known as the measurement line and the datum line\,---\,corresponds to the official length of the track.}
and the inclination of the track.
Our expressions are accurate analogies for the common geometry of modern $250$\,-metre velodromes~(Mehdi Kordi, {\it pers.~comm.}, 2020).
We proceed to formulate an expression for power expended against dissipative forces, which we examine for both the constant-cadence and constant-power cases.
We examine their empirical adequacy, and conclude by discussing the results.
In the appendices, we consider, {\it a posteriori}, changes in the kinetic and potential energy, for both the constant-cadence and constant-power cases, as well as the explicite measurements: force and cadence.
\section{Track}
\label{sec:Formulation}
\subsection{Black-line parameterization}
\label{sub:Track}
\begin{figure}[h]
\centering
\includegraphics[scale=0.35]{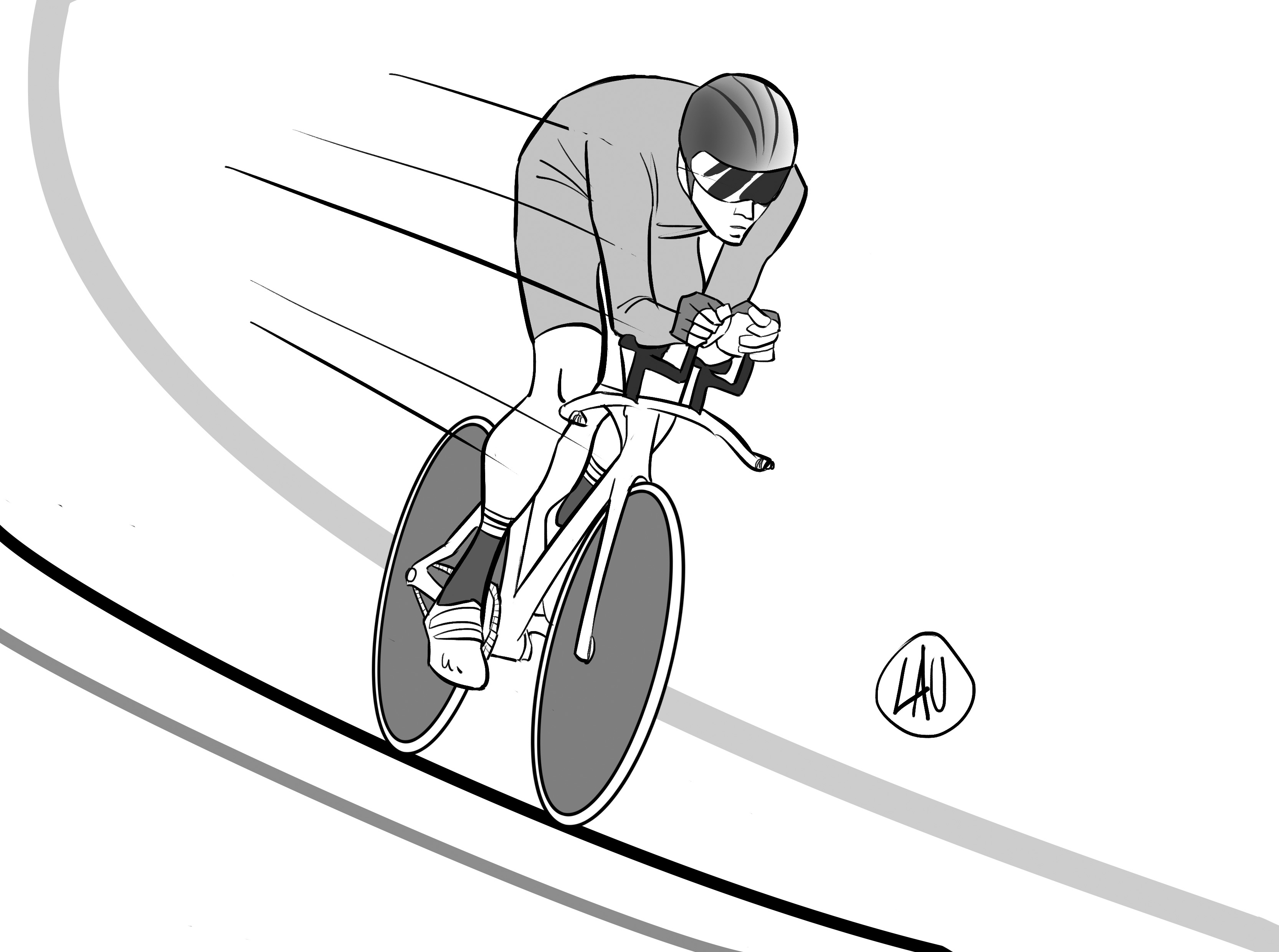}
\caption{\small  A constant aerodynamic position along the black line}
\label{fig:FigBlackLine}
\end{figure}

\begin{figure}[h]
\centering
\includegraphics[scale=0.5]{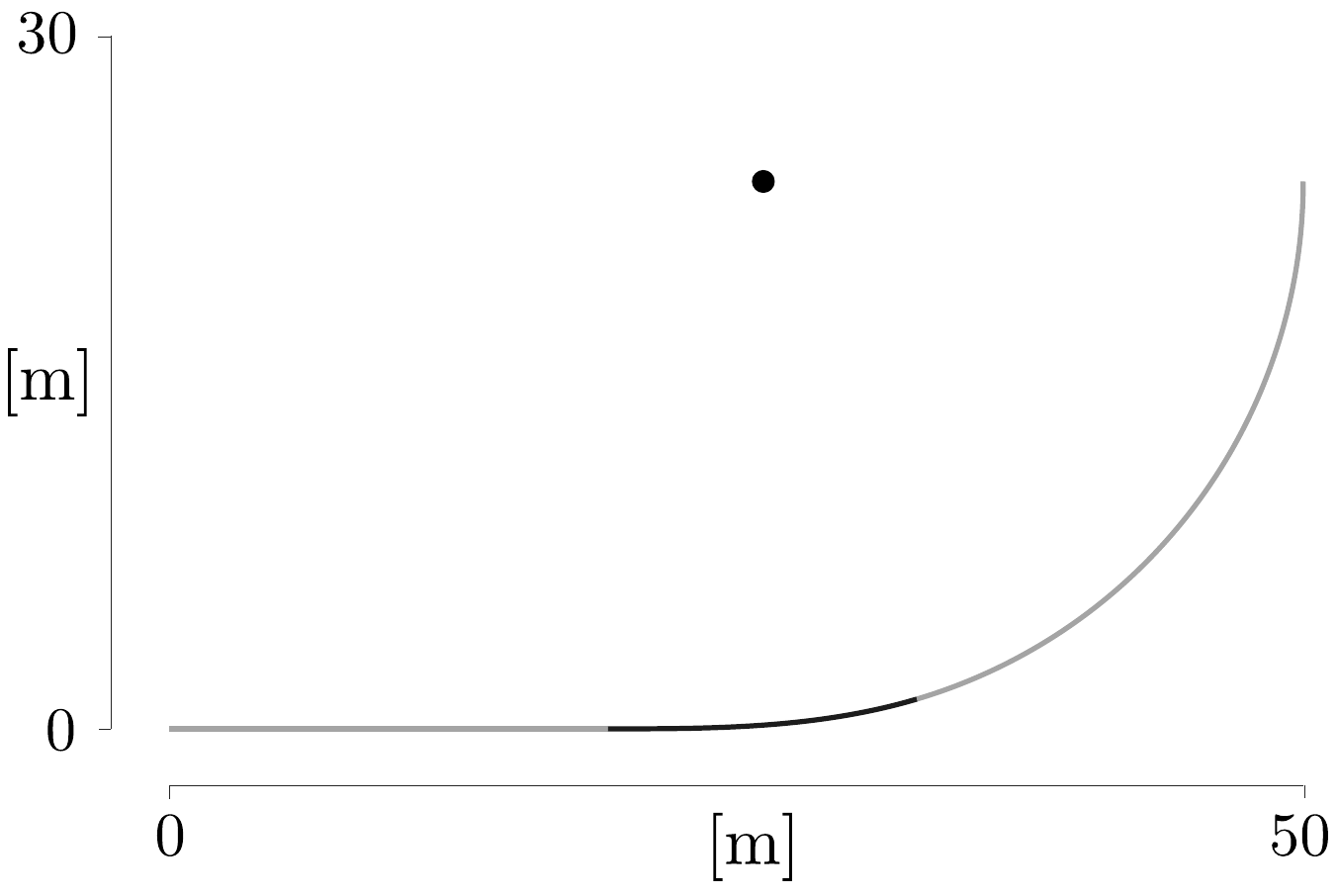}
\caption{\small  A quarter of the black line for a $250$\,-metre track}
\label{fig:FigTrack}
\end{figure}

To model the required power for an individual pursuit of a cyclist who follows the black line, in a constant aerodynamic position, as illustrated in Figure~\ref{fig:FigBlackLine}, we define this line by three parameters.
\begin{itemize}
\item[--] $L_s$\,: the half-length of the straight
\item[--] $L_t$\,: the length of the transition curve between the straight and the circular arc
\item[--] $L_a$\,: the half-length of the circular arc
\end{itemize}
The length of the track is $S=4(L_s+L_t+L_a)$\,.
In Figure~\ref{fig:FigTrack}, we show a quarter of a black line for $L_s=19$\,m\,, $L_t=13.5$\,m and $L_a=30$\,m\,, which results in $S=250$\,m\,. 
This curve has continuous derivative up to order two; it is a $C^2$ curve,  whose curvature is continuous.

To formulate, in Cartesian coordinates, the curve shown in Figure~\ref{fig:FigTrack}, we consider the following.
\begin{itemize}
\item[--] The straight,
\begin{equation*}
y_1=0\,,\qquad0\leqslant x\leqslant a\,,	
\end{equation*}
shown in gray, where $a:=L_s$\,.
\item[--] The transition, shown in black\,---\,following a standard design practice\,---\,we take to be an Euler spiral, which can be parameterized by Fresnel integrals,
\begin{equation*}
x_2(\varsigma)=a+\sqrt{\frac{2}{A}}\int\limits_0^{\varsigma\sqrt{\!\frac{A}{2}}}\!\!\!\!\cos\!\left(x^2\right)\,{\rm d}x
\end{equation*}
and
\begin{equation*}
y_2(\varsigma)=\sqrt{\frac{2}{A}}\int\limits_0^{\varsigma\sqrt{\!\frac{A}{2}}}\!\!\!\!\sin\!\left(x^2\right)\,{\rm d}x\,,
\end{equation*}
with $A>0$ to be determined; herein, $\varsigma$ is a curve parameter.
Since the arclength differential,~${\rm d}s$\,, is such that
\begin{align*}
{\rm d}s&=\sqrt{x_2'(\varsigma)^2+y_2'(\varsigma)^2}\,{\rm d}\varsigma\\
&=\sqrt{\cos^2\left(\dfrac{A\varsigma^2}{2}\right)+\sin^2\left(\dfrac{A\varsigma^2}{2}\right)}\,{\rm d}\varsigma\\
&={\rm d}\varsigma\,,
\end{align*}
we write the transition curve as
\begin{equation*}
(x_2(s),y_2(s)), \quad 0\leqslant s\leqslant b:=L_t\,.
\end{equation*}
\item[--] The circular arc, shown in gray, whose centre is $(c_1,c_2)$ and whose radius is $R$\,, with $c_1$\,, $c_2$ and $R$  to be determined.
Since its arclength is specified to be $c:=L_a,$ we may parameterize the quarter circle by
\begin{equation}
\label{eq:x3}
x_3(\theta)=c_1+R\cos(\theta)
\end{equation}
and
\begin{equation}
\label{eq:y3}
y_3(\theta)=c_2+R\sin(\theta)\,,
\end{equation}
where $-\theta_0\leqslant\theta\leqslant 0$\,, for $\theta_0:=c/R$\,.
The centre of the circle is shown as a black dot in Figure~\ref{fig:FigTrack}.
\end{itemize}

We wish to connect these three curve segments so that the resulting global curve is continuous along with its first and second derivatives.
This ensures that the curvature of the track is also continuous.

To do so, let us consider the connection between the straight and the Euler spiral.
Herein, $x_2(0)=a$ and $y_2(0)=0$\,, so the spiral connects continuously to the end of the straight at $(a,0)$\,.
Also, at $(a,0)$\,,
\begin{equation*}
\frac{{\rm d}y}{{\rm d}x}=\frac{y_2'(0)}{x_2'(0)}=\frac{0}{1}=0\,,
\end{equation*}
which matches the derivative of the straight line.
Furthermore, the second derivatives match, since
\begin{equation*}
\frac{{\rm d}^2y}{{\rm d}x^2}=\frac{y''_2(0)x_2'(0)-y'_2(0)x_2''(0)}{(x_2'(0))^2}=0\,,
\end{equation*}
which follows, for any $A>0$\,, from
\begin{equation}
\label{eq:FirstDer}
x_2'(\varsigma)=\cos^2\left(\dfrac{A\,\varsigma^2}{2}\right)\,, \quad y_2'(\varsigma)=\sin^2\left(\dfrac{A\,\varsigma^2}{2}\right)
\end{equation}
and
\begin{equation*}
x_2''(\varsigma)=-A\,\varsigma\sin\left(\dfrac{A\,\varsigma^2}{2}\right)\,, \quad y_2''(\varsigma)=A\,\varsigma\cos\left(\dfrac{A\,\varsigma^2}{2}\right)\,.
\end{equation*}
Let us consider the connection between the Euler spiral and the arc of the circle. 
In order that these connect continuously,
\begin{equation*}
\big(x_2(b),y_2(b)\big)=\big(x_3(-\theta_0),y_3(-\theta_0)\big)\,,
\end{equation*}
we require
\begin{equation}
\label{eq:Cont1}
x_2(b)=c_1+R\cos(\theta_0)\,\,\iff\,\,c_1=x_2(b)-R\cos\!\left(\dfrac{c}{R}\right)
\end{equation}
and
\begin{equation}
\label{eq:Cont2}
y_2(b)=c_2-R\sin(\theta_0)\,\,\iff\,\, c_2=y_2(b)+R\sin\!\left(\dfrac{c}{R}\right)\,.
\end{equation}
For the tangents to connect continuously, we invoke expression~(\ref{eq:FirstDer}) to write
\begin{equation*}
(x_2'(b),y_2'(b))=\left(\cos\left(\dfrac{A\,b^2}{2}\right),\,\sin\left(\dfrac{A\,b^2}{2}\right)\right)\,.
\end{equation*}
Following expressions~(\ref{eq:x3}) and (\ref{eq:y3}), we obtain
\begin{equation*}
\big(x_3'(-\theta_0),y_3'(-\theta_0)\big)=\big(R\sin(\theta_0),R\cos(\theta_0)\big)\,,
\end{equation*}
respectively.
Matching the unit tangent vectors results in
\begin{equation}
\label{eq:tangents}
\cos\left(\dfrac{A\,b^2}{2}\right)=\sin\!\left(\dfrac{c}{R}\right)\,,\quad \sin\left(\dfrac{A\,b^2}{2}\right)=\cos\!\left(\dfrac{c}{R}\right)\,.
\end{equation}
For the second derivative, it is equivalent\,---\,and easier\,---\,to match the curvature.
For the Euler spiral,
\begin{align*}
\kappa_2(s)&=\frac{x_2'(s)y_2''(s)-y_2'(s)x_2''(s)}
{\Big(\big(x_2'(s)\big)^2+\big(y_2'(s)\big)^2\Big)^{\frac{3}{2}}}\\
&=A\,s\cos^2\left(\dfrac{A\,s^2}{2}\right)+A\,s\sin^2\left(\dfrac{A\,s^2}{2}\right)\\
&=A\,s\,,
\end{align*}
which is indeed the defining characteristic of an Euler spiral: the curvature grows linearly in the arclength.
Hence, to match the curvature of the circle at the connection, we require
\begin{equation*}
A\,b=\frac{1}{R} \,\,\iff\,\,A=\frac{1}{b\,R}\,.
\end{equation*}
Substituting this value of $A$ in equations~(\ref{eq:tangents}), we obtain
\begin{align*}
\cos\!\left(\dfrac{b}{2R}\right)&=\sin\!\left(\dfrac{c}{R}\right)\,,\quad \sin\!\left(\dfrac{b}{2R}\right)=\cos\!\left(\dfrac{c}{R}\right)\\
&\iff\dfrac{b}{2R}=\dfrac{\pi}{2}-\dfrac{c}{R}\\
&\iff R=\frac{b+2c}{\pi}.
\end{align*}
It follows that
\begin{equation*}
A=\frac{1}{b\,R}=\frac{\pi}{b\,(b+2c)}\,;
\end{equation*}
hence, the continuity condition stated in expressions~(\ref{eq:Cont1}) and (\ref{eq:Cont2}) determines the centre of the circle,~$(c_1,c_2)$\,.

\begin{figure}[h]
\centering
\includegraphics[scale=0.5]{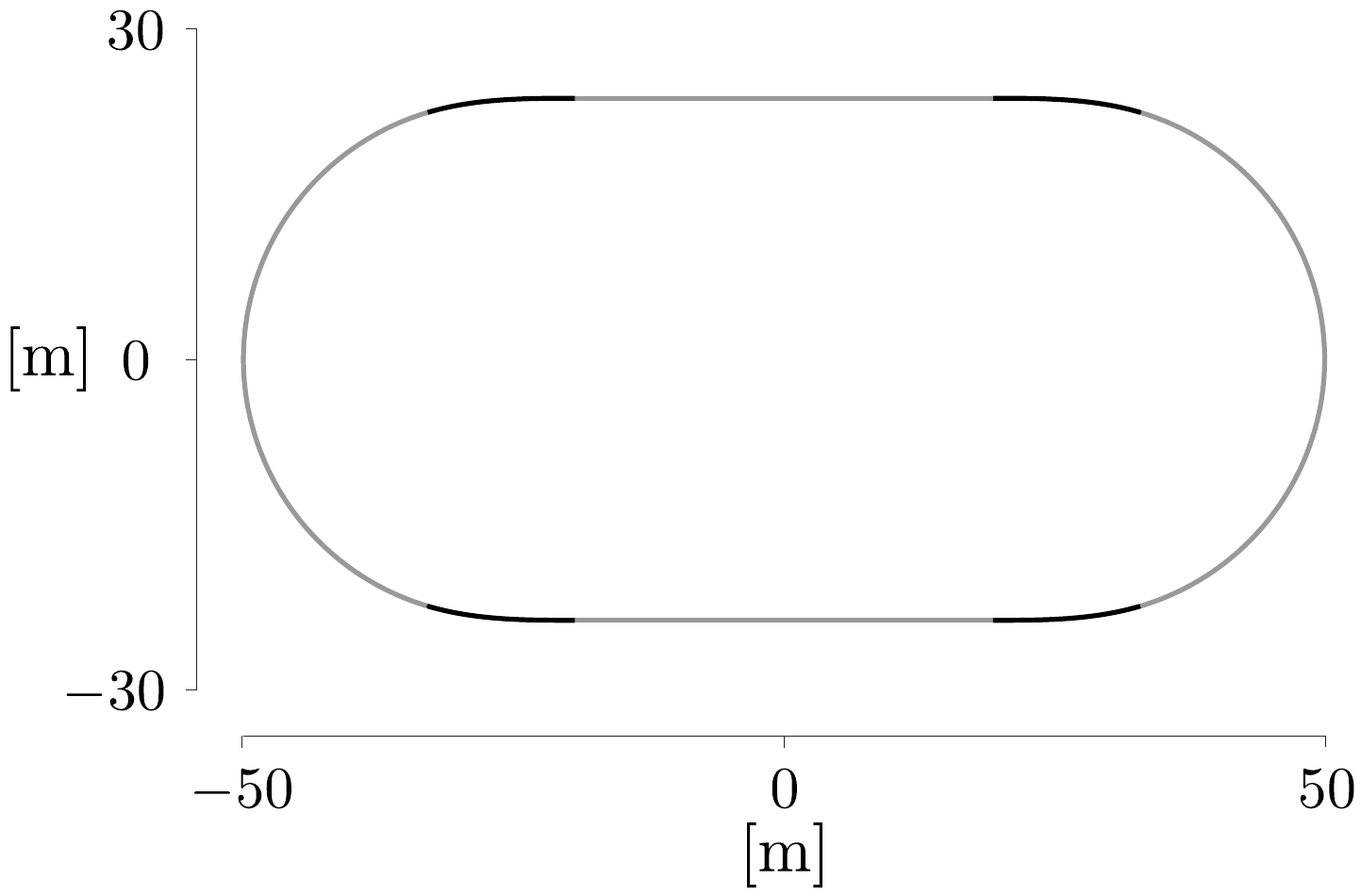}
\caption{\small Black line of $250$\,-metre track}
\label{fig:FigComplete}
\end{figure}

\begin{figure}[h]
\centering
\includegraphics[scale=0.5]{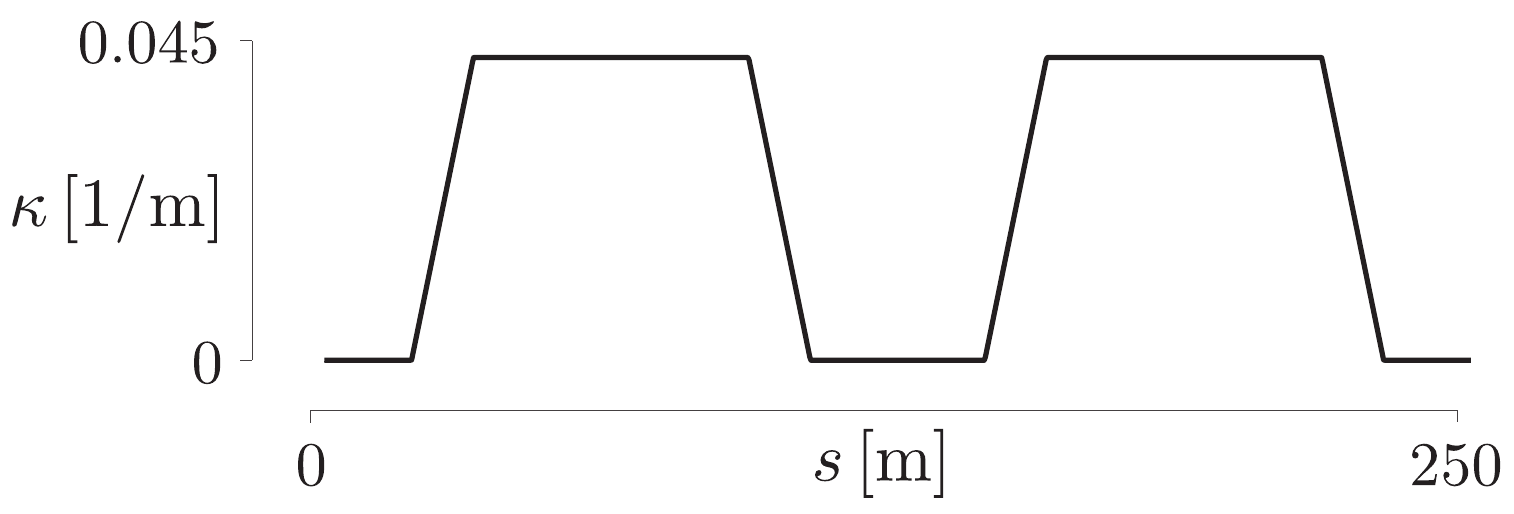}
\caption{\small Curvature of the black line,~$\kappa$\,, as a function of distance,~$s$\,, with a linear transition between the zero curvature of the straight and the $1/R$ curvature of the circular arc}
\label{fig:FigCurvature}
\end{figure}

For the case shown in Figure~\ref{fig:FigTrack}, the numerical values are~$A=3.1661\times10^{-3}$\,m${}^{-2}$, $R=23.3958$\,m\,, $c_1=25.7313$\,m and $c_2=23.7194$\,m\,.
The complete track\,---\,with its centre at the origin\,,~$(0,0)$\,---\,is shown in Figure~\ref{fig:FigComplete}.
The corresponding curvature is shown in Figure~\ref{fig:FigCurvature}.
Note that  the curvature transitions linearly from the constant value of straight,~$\kappa=0$\,, to the constant value of the circular arc,~$\kappa=1/R$\,.
\subsection{Track-inclination angle}
\begin{figure}[h]
\centering
\includegraphics[scale=0.5]{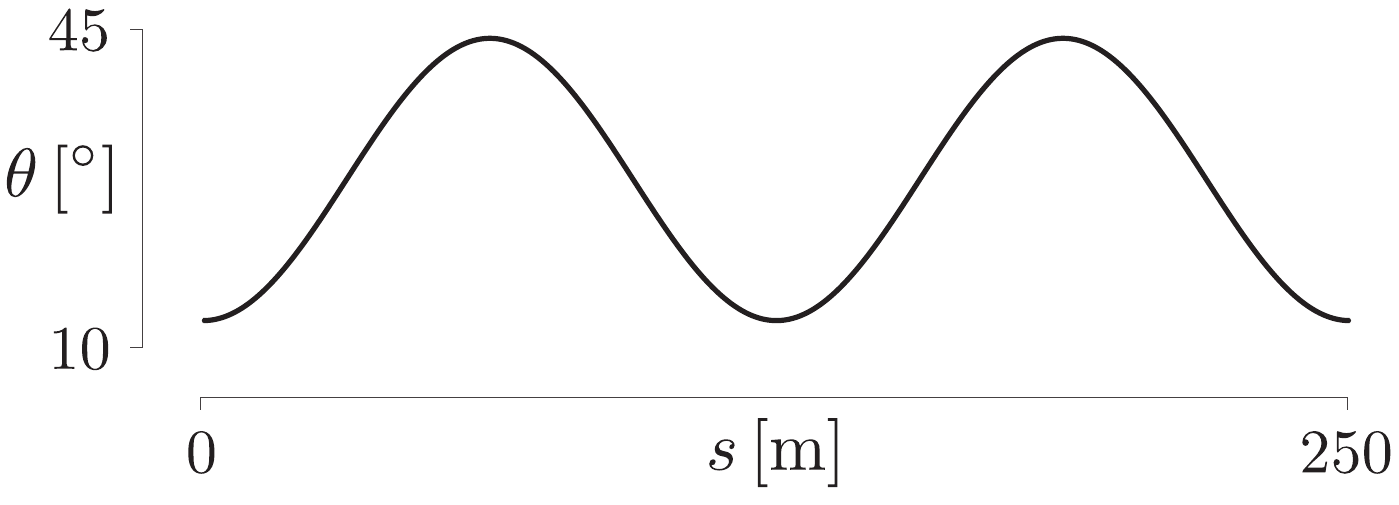}
\caption{\small Track inclination,~$\theta$\,, as a function of the black-line distance,~$s$}
\label{fig:FigAngle}
\end{figure}
There are many possibilities to model the track inclination angle.
We choose a trigonometric formula in terms of arclength, which is a good analogy of an actual $250$\,-metre velodrome.
The minimum inclination of $13^\circ$ corresponds to the midpoint of the straight, and the maximum of $44^\circ$ to the apex of the circular arc.
For a track of length $S$\,,
\begin{equation}
\label{eq:theta}
\theta(s)=28.5-15.5\cos\!\left(\frac{4\pi}{S}s\right)\,;
\end{equation}
$s=0$ refers to the midpoint of the lower straight, in Figure~\ref{fig:FigComplete}, and the track is oriented in the counterclockwise direction.
Figure \ref{fig:FigAngle} shows this inclination for $S=250$\,m\,.
\section{Instantaneous power}
\label{sec:InstPower}
\begin{figure}[h]
\centering
\includegraphics[scale=0.8]{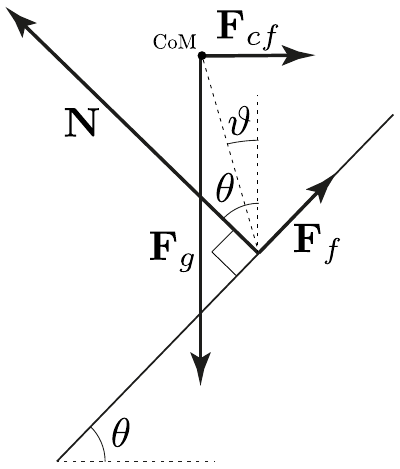}
\caption{\small Force diagram}
\label{fig:FigCentFric}
\end{figure}

A mathematical model to account for the power required to propel a bicycle is based on \citep[e.g.,][]{DSSbici1}
\begin{equation}
\label{eq:BikePower}
P=F\,V\,,
\end{equation}
where $F$ stands for the magnitude of forces opposing the motion and $V$ for speed.
Herein, we model the rider as undergoing instantaneous circular motion, in rotational equilibrium about the line of contact of the tires with the ground.
Following \citet[Section~2]{SSSbici3}, in accordance with Figure~\ref{fig:FigCentFric}, along the black line of a velodrome, in windless conditions,
\begin{subequations}
\label{eq:power}
\begin{align}
\nonumber P&=\\
&\dfrac{1}{1-\lambda}\,\,\Bigg\{\label{eq:modelO}\\
&\left.\left.\Bigg({\rm C_{rr}}\underbrace{\overbrace{\,m\,g\,}^{F_g}(\sin\theta\tan\vartheta+\cos\theta)}_N\cos\theta
+{\rm C_{sr}}\Bigg|\underbrace{\overbrace{\,m\,g\,}^{F_g}\frac{\sin(\theta-\vartheta)}{\cos\vartheta}}_{F_f}\Bigg|\sin\theta\Bigg)\,v
\right.\right.\label{eq:modelB}\\
&+\,\,\tfrac{1}{2}\,{\rm C_{d}A}\,\rho\,V^3\Bigg\}\label{eq:modelC}\,,
\end{align}	
\end{subequations}
where $m$ is the mass of the cyclist and the bicycle, $g$ is the acceleration due to gravity, $\theta$ is the track-inclination angle, $\vartheta$ is the bicycle-cyclist lean angle, $\rm C_{rr}$ is the rolling-resistance coefficient, $\rm C_{sr}$ is the coefficient of the lateral friction, $\rm C_{d}A$ is the air-resistance coefficient, $\rho$ is the air density, $\lambda$ is the drivetrain-resistance coefficient.
Herein, $v$ is the speed at which the contact point of the rotating wheels moves along the track \citep[Appendix~B]{DSSbici1}, which we commonly consider as coinciding with the black-line speed.
$V$ is the centre-of-mass speed.
Since the lateral friction is a dissipative force, it does negative work, and the work done against it\,---\,as well as the power\,---\,are positive.
For this reason, in expression~(\ref{eq:modelB}), we consider the magnitude,~$\big|{\,\,}\big|$\,.

For reasons discussed by \citet[Appendix~B.2]{DSSbici2}, in expression~(\ref{eq:power}), we assume the steadiness of effort, which\,---\,following an initial acceleration\,---\,is consistent with a steady pace of an individual pursuit, as presented in Section~\ref{sec:Adequacy}, below.
Formally, this assumption corresponds to setting the acceleration,~$a$\,, to zero in \citet[expression~(1)]{SSSbici3}.
Herein, the acceleration refers to the change of the centre-of-mass speed.
This speed is nearly constant if the power is constant, which can be viewed as a quantification of the cyclist's effort.
In other words, the force\,---\,and hence the power required to accelerate the bicycle-cyclist system\,---\,is associated mainly with the change of the centre-of-mass speed, not with the change of the black-line speed.

To gain an insight into expression~(\ref{eq:power}), let us consider a few special cases.
If $\theta=\vartheta=0$\,,
\begin{equation}
\label{eq:straight}
P=\underbrace{\dfrac{{\rm C_{rr}}\,m\,g+\tfrac{1}{2}\,{\rm C_{d}A}\,\rho\,V^2}{1-\lambda}}_F\,V\,,
\end{equation}
where\,---\,as expected for a flat, straight road\,---\,$v\equiv V$\,.
Also, on a velodrome, along the straights, $\vartheta=0$\, and expression~(\ref{eq:modelB}) becomes
\begin{equation*}
\left({\rm C_{rr}}\,m\,g\,\cos^2\theta
+{\rm C_{sr}}\,m\,g\,\sin^2\theta\right)\,V\,.
\end{equation*}
If, along the curves, $\vartheta=\theta$\,, the second summand of expression~(\ref{eq:modelB}) is zero, as expected.

Let us return to expression~(\ref{eq:power}).
Therein, $\theta$ is given by expression~(\ref{eq:theta}).
The lean angle is \citep[Appendix~A]{SSSbici3}
\begin{equation}
\label{eq:LeanAngle}
\vartheta=\arctan\dfrac{V^2}{g\,r_{\rm\scriptscriptstyle CoM}}\,,
\end{equation}
where $r_{\rm\scriptscriptstyle CoM}$ is the centre-of-mass radius, and\,---\,along the curves, at any instant\,---\,the centre-of-mass speed is
\begin{equation}
\label{eq:vV}
V=v\,\dfrac{\overbrace{(R-h\sin\vartheta)}^{\displaystyle r_{\rm\scriptscriptstyle CoM}}}{R}
=v\,\left(1-\dfrac{h\,\sin\vartheta}{R}\right)\,,
\end{equation}
where $R$ is the radius discussed in Section~\ref{sub:Track} and $h$ is the centre-of-mass height.
Along the straights, the  black-line speed is equivalent to the centre-of-mass speed, $v=V$\,.
As expected, $V=v$ if $h=0$\,, $\vartheta=0$ or $R=\infty$\,.

Invoking expressions~(\ref{eq:LeanAngle}) and (\ref{eq:vV}), we neglect the vertical variation of the centre of mass and, hence, assume that the centre-of-mass trajectory is contained in a horizontal plane, where\,---\,in accordance with the track geometry\,---\,this plane is parallel to the plane that contains the black line.
Accounting for the vertical motion of the centre of mass would mean allowing for a nonhorizontal centripetal force and including the work done in raising the centre of mass.
\section{Numerical examples}
\label{sec:NumEx}
\subsection{Model-parameter values}
\label{sub:ModPar}
For expressions~(\ref{eq:power}), (\ref{eq:LeanAngle}) and (\ref{eq:vV}), we consider a velodrome discussed in Section~\ref{sec:Formulation}, and let $R=23.3958$\,m\,.
For the bicycle-cyclist system, we assume, $h=1.2$\,m\,, $m=84$\,kg\,, ${\rm C_{d}A}=0.2$\,m${}^2$\,, ${\rm C_{rr}}=0.002$\,, ${\rm C_{sr}}=0.003$ and $\lambda=0.02$\,.
For the external conditions, $g=9.81$\,m/s${}^2$ and $\rho=1.225$\,kg/m${}^3$\,.
\subsection{Constant cadence}
\label{sub:ConstCad}
Let the black-line speed be constant,~$v=16.7$\,m/s\,, which is tantamount to the constancy of cadence.
As discussed in Section~\ref{sec:InstPower}, the assumption of a constant black-line speed means neglecting the acceleration of the centre of mass.

\begin{figure}[h]
\centering
\includegraphics[scale=0.5]{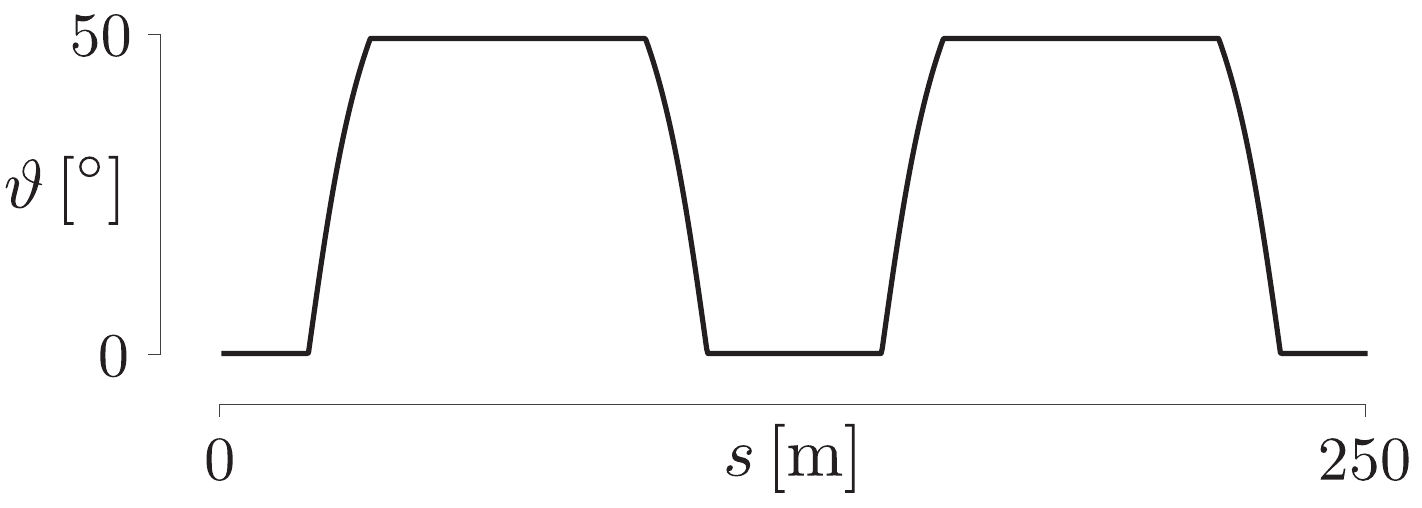}
\caption{\small Lean angle,~$\vartheta$\,, as a function of the black-line distance,~$s$\,, for constant cadence}
\label{fig:FigLeanAngle}
\end{figure}

\begin{figure}[h]
\centering
\includegraphics[scale=0.5]{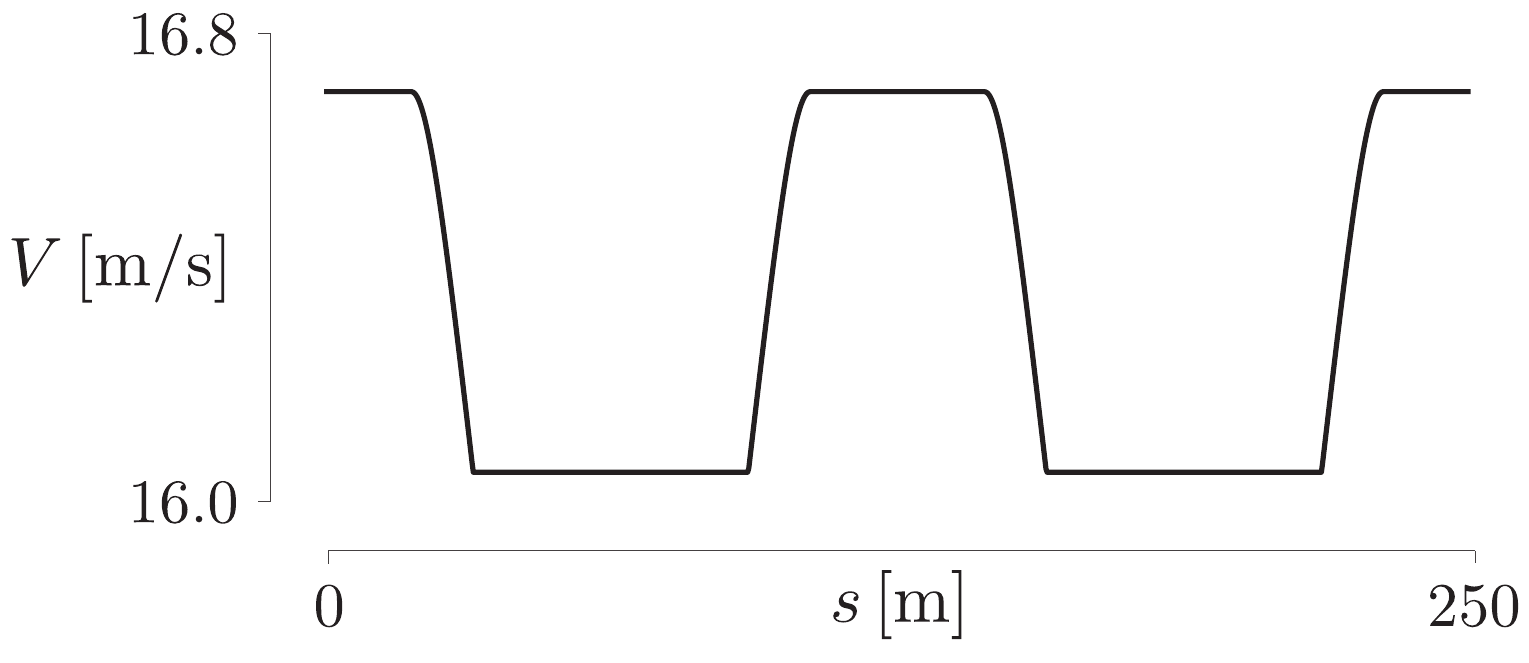}
\caption{\small Centre-of-mass speed,~$V$\,, as a function of the black-line distance,~$s$\,, for constant cadence}
\label{fig:FigCoMSpeed}
\end{figure}

The lean angle and the centre-of-mass speed, as functions of distance\,---\,obtained by numerically and simultaneously solving equations~(\ref{eq:LeanAngle}) and (\ref{eq:vV}), at each point of a discretized model of  the track\,---\,are shown in Figures~\ref{fig:FigLeanAngle} and \ref{fig:FigCoMSpeed}, respectively.
The average centre-of-mass speed, per lap is~$\overline V=16.3329$\,m/s\,.
Changes of $V$\,, shown in Figure~\ref{fig:FigCoMSpeed}, result from the lean angle.
Along the straights, $\vartheta=0\implies V=v$\,.
Along the curves, since $\vartheta\neq0$\,, the centre-of-mass travels along a shorter path; hence, $V<v$\,.
Thus, assuming a constant black-line speed implies a variable centre-of-mass speed and, hence, an acceleration and deceleration, even though ${\rm d}V/{\rm d}t$\,, where $t$ stands for time, is not included explicitly in expression~(\ref{eq:power}).
Examining Figure~\ref{fig:FigCoMSpeed}, we conclude that ${\rm d}V/{\rm d}t\neq0$ along the transition curves only.

\begin{figure}[h]
\centering
\includegraphics[scale=0.5]{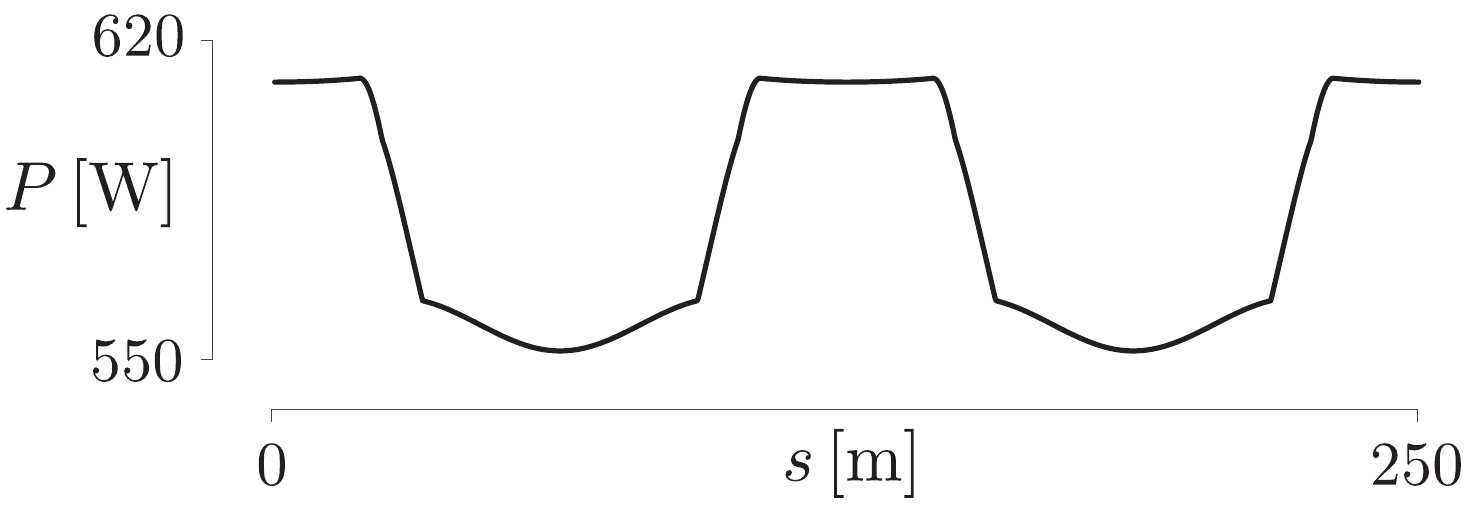}
\caption{\small Power,~$P$\,, as a function of the black-line distance,~$s$\,, for constant cadence}
\label{fig:FigPower}
\end{figure}

The power\,---\,obtained by evaluating expression~(\ref{eq:power}), at each point along the track\,---\,is shown in Figure~\ref{fig:FigPower}.
The average power, per lap, is $\overline P=580.5941$\,W\,.
Since the black-line speed is constant, this is both the arclength average and the temporal average.
\begin{figure}[h]
\centering
\includegraphics[scale=0.5]{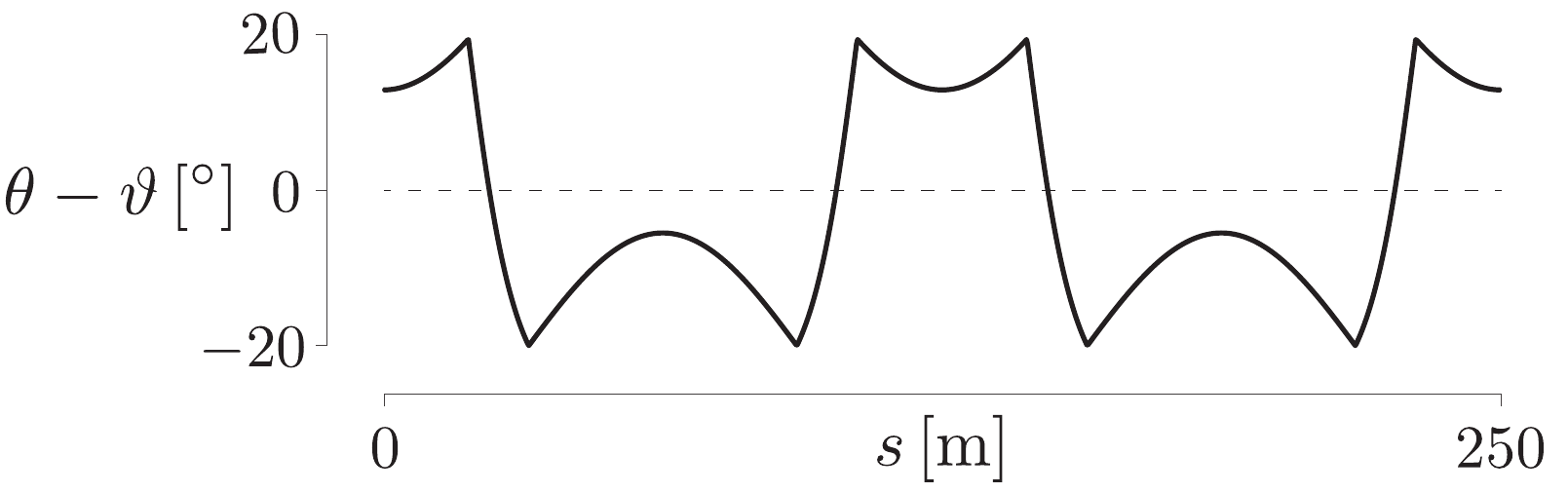}
\caption{\small $\theta-\vartheta$\,, as a function of the black-line distance,~$s$\,, for constant cadence}
\label{fig:FigAngleDiff}
\end{figure}
\begin{figure}[h]
\centering
\includegraphics[scale=0.5]{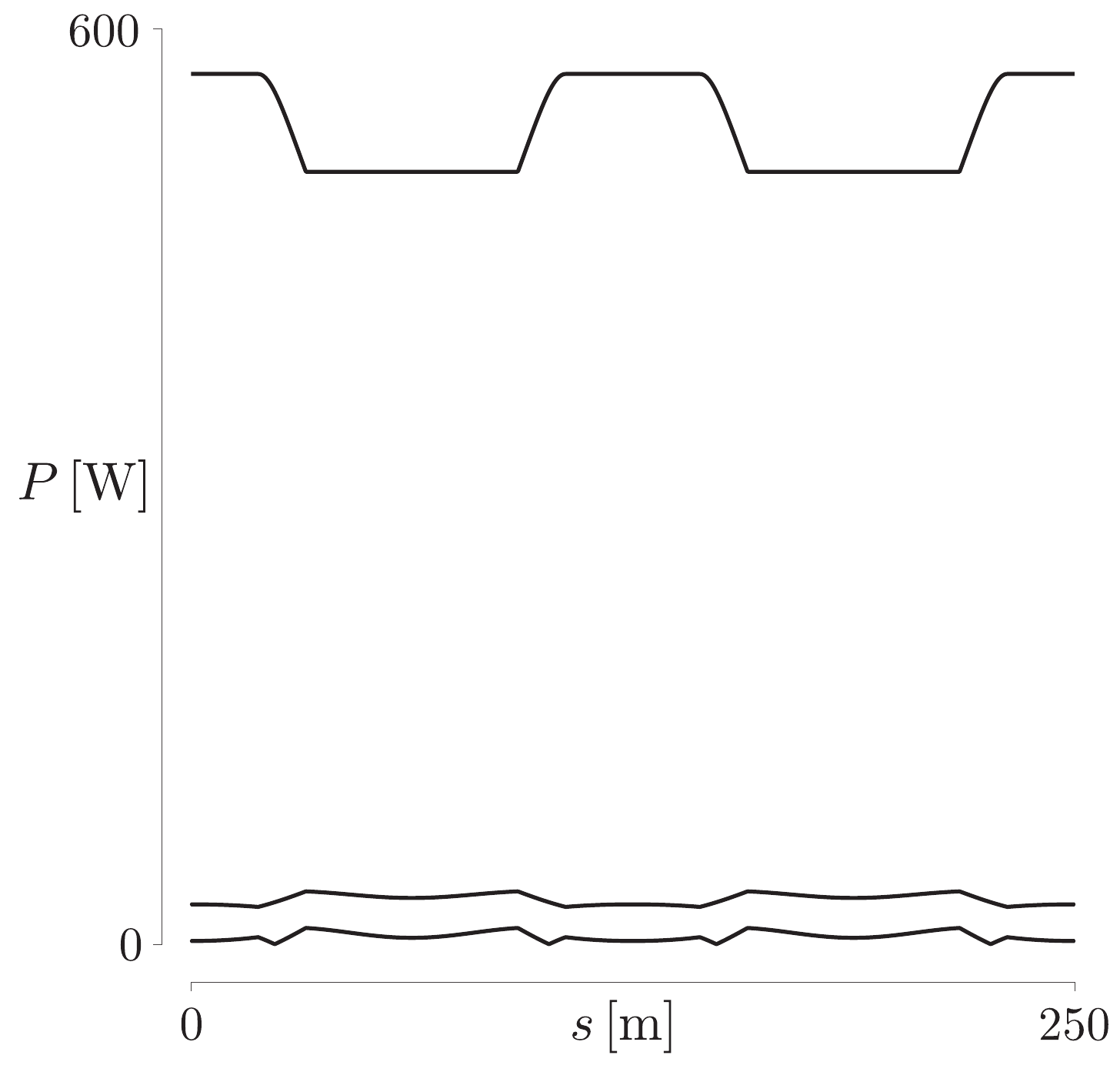}
\caption{\small Power to overcome air resistance, rolling resistance and lateral friction}
\label{fig:FigPowerSummands}
\end{figure}

Examining Figure~\ref{fig:FigPower}, we see the decrease of power required to maintain the same black-line speed along the curve.
This is due to both the decrease of the centre-of-mass speed, which results in a smaller value of term~(\ref{eq:modelC}), and the decrease of a difference between the track-inclination angle and the lean angle, shown in Figure~\ref{fig:FigAngleDiff}, which results in a smaller value of the second summand of term~(\ref{eq:modelB}).

The argument presented in the previous paragraph leads to the following conjecture.
The most efficient track is circular with $\theta=\vartheta$\,, which would correspond to the dashed line in Figure~\ref{fig:FigAngleDiff}.
However, this is not possible, since\,---\,according to the regulations of the Union Cycliste Internationale\,---\,the inner edge of the track shall consist of two curves connected by two parallel straight lines.
Hence, the optimization is constrained by the length of the straights.

Examining Figure~\ref{fig:FigPowerSummands}, where\,---\,in accordance with expression~(\ref{eq:power})\,---\,we distinguish among the power used to overcome the air resistance, the rolling resistance and the lateral friction, we can quantify their effects.
The first has the most effect; the last has the least effect, and is zero at points for which $\theta=\vartheta$\,, which corresponds to the zero crossings in Figure~\ref{fig:FigAngleDiff}.

Let us comment on potential simplifications of a model.
If we assume a straight flat course\,---\,which is tantamount to neglecting the lean and inclination angles\,---\,we obtain, following expression~(\ref{eq:straight}), $\overline P\approx 610$\,W\,.
If we consider an oval track but ignore the transitions and assume that the straights are flat and the semicircular segments, whose radius is $23$\,m\,, have a constant inclination of $43^\circ$, we obtain \citep[expression~(13)]{SSSbici3} $\overline P\approx 563$\,W\,.
In both cases, there is a significant discrepancy with the power obtained from the model discussed herein,~$\overline P=573.6080$\,W\,.

To conclude this section, let us calculate the work per lap corresponding to the model discussed herein.
The work performed during a time interval, $t_2-t_1$\,, is
\begin{equation*}
W=\int\limits_{t_1}^{t_2}\!P\,{\rm d}t
=\dfrac{1}{v}\int\limits_{s_1}^{s_2}\!P\!\underbrace{\,v\,{\rm d}t}_{{\rm d}s\,}\,,	
\end{equation*}
where the black-line speed,~$v$\,, is constant and, hence, ${\rm d}s$ is an arclength distance along the black line.
Considering the average power per lap, we write
\begin{equation}
\label{eq:WorkConstCad}
W=\underbrace{\,\dfrac{S}{v}\,}_{t_\circlearrowleft}\,\underbrace{\dfrac{\int\limits_0^S\!P\,{\rm d}s}{S}}_{\overline P}=\overline P\,t_\circlearrowleft\,.
\end{equation}
Given $\overline P=580.5941$\,W and $t_\circlearrowleft=14.9701$\,s\,, we obtain $W=8691.5284$\,J\,.
\subsection{Constant power}
\label{sub:ConstPower}
Let us solve numerically the system of nonlinear equations given by  expressions~(\ref{eq:power}),  (\ref{eq:LeanAngle}) and (\ref{eq:vV}), to find the lean angle as well as both speeds, $v$ and $V$\,, at each point of a discretized model of the track\,, under the assumption of  constant power.
In accordance with a discussion in Section~\ref{sec:InstPower}, such an assumption is more consistent with the steadiness of effort than the assumption of a constant cadence examined in Section~\ref{sub:ConstCad}.

As in Section~\ref{sub:ConstCad}, we let $R=23.3958$\,m\,, $h=1.2$\,m\,, $m=84$\,kg\,, ${\rm C_{d}A}=0.2$\,m${}^2$\,, ${\rm C_{rr}}=0.002$\,, ${\rm C_{sr}}=0.003$\,, $\lambda=0.02$\,, $g=9.81$\,m/s${}^2$ and $\rho=1.225$\,kg/m${}^3$\,.
However, in contrast to Section~\ref{sub:ConstCad}, we allow the black-line speed to vary, and set the power to be the average obtained in that section, $P=580.5941$\,W\,.

Stating expression~(\ref{eq:vV}), as
\begin{equation*}
v=V\dfrac{R}{R-h\sin\vartheta}\,,
\end{equation*}
we write expression~(\ref{eq:power}) as
\begin{align}
\label{eq:PConst}
P&=\\
\nonumber&\dfrac{V}{1-\lambda}\,\,\Bigg\{\\
\nonumber&\left.\left.\Bigg({\rm C_{rr}}\,m\,g\,(\sin\theta\tan\vartheta+\cos\theta)\cos\theta
+{\rm C_{sr}}\Bigg|\,m\,g\,\frac{\sin(\theta-\vartheta)}{\cos\vartheta}\Bigg|\sin\theta\Bigg)\,\dfrac{R}{R-h\sin\vartheta}
\right.\right.\\
\nonumber&+\,\,\tfrac{1}{2}\,{\rm C_{d}A}\,\rho\,V^2\Bigg\}\,,
\end{align}
and expression~(\ref{eq:LeanAngle}) as
\begin{equation}
\label{eq:Vvar}
\vartheta=\arctan\dfrac{V^2}{g\,(R-h\sin\vartheta)}\,,
\end{equation}
which\,---\,given $g$\,, $R$ and $h$\,---\,can be solved for $V$ as a function of~$\vartheta$\,.
Inserting that solution in expression~(\ref{eq:PConst}), we obtain an equation whose only unknown is~$\vartheta$\,.

The difference of the lean angle\,---\,between the case of a constant cadence and a constant power is so small that there is no need to plot it; Figure~\ref{fig:FigLeanAngle} illustrates it accurately.
The same is true for the difference between the track-inclination angle and the lean angle, illustrated in Figure~\ref{fig:FigAngleDiff}, as well as for the dominant effect of the air resistance, illustrated in Figure~\ref{fig:FigPowerSummands}.

\begin{figure}[h]
\centering
\includegraphics[scale=0.5]{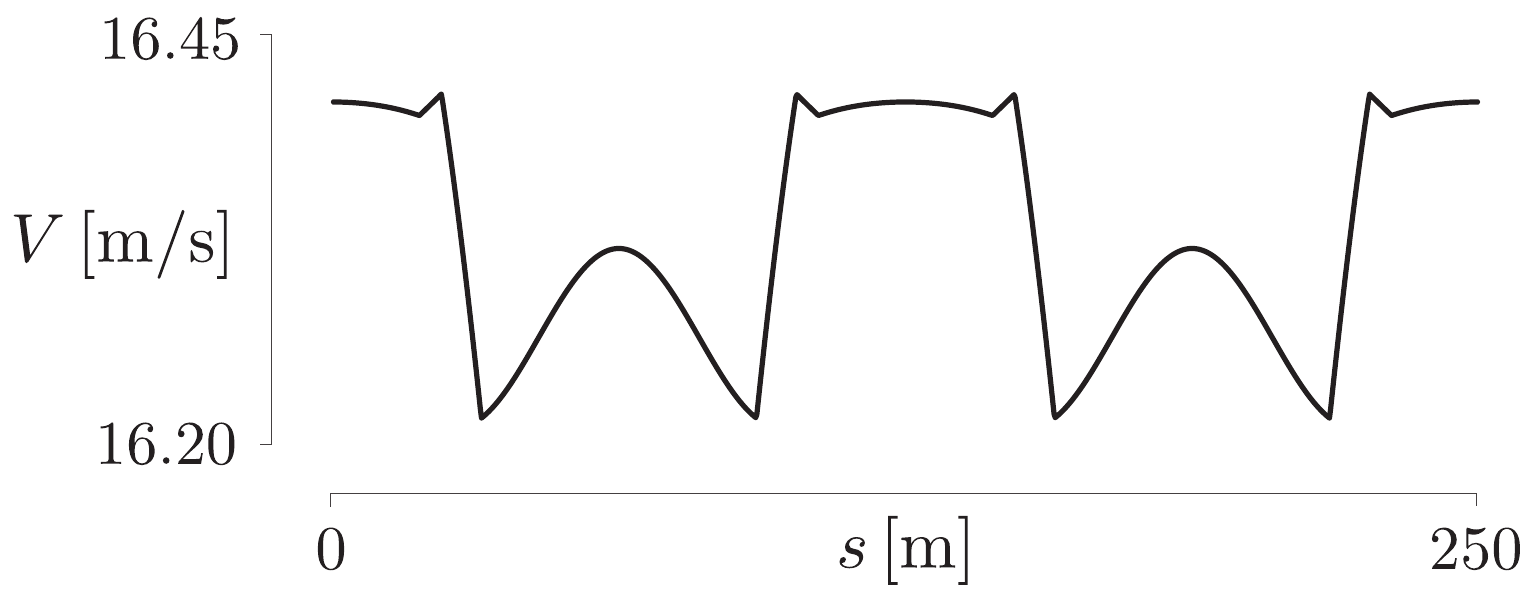}
\caption{\small Centre-of-mass speed,~$V$\,, as a function of the black-line distance,~$s$\,, for constant power}
\label{fig:FigCoMSpeed2}
\end{figure}

The resulting values of $V$ are shown in Figure~\ref{fig:FigCoMSpeed2}.
As expected, in view of the dominant effect of the air resistance, a constancy of $P$ entails only small variations in $V$\,.
In comparison to the case discussed in Section~\ref{sub:ConstCad}, the case in question entails lesser accelerations and decelerations of the centre of mass\,---\,note the difference of vertical scale between Figures~\ref{fig:FigCoMSpeed} and \ref{fig:FigCoMSpeed2}\,---\,but the changes of speed are not limited to the transition curves.
Even though such changes are not included explicitly in expression~(\ref{eq:power}), a portion of the given power may be accounted for by $m\,V\,{\rm d}V/{\rm d}t$\,, which is associated with accelerations and decelerations.
The amount of this portion can be estimated {\it a posteriori}.

Since
\begin{equation}
\label{eq:dK}
m\,V\,\dfrac{{\rm d}V}{{\rm d}t}=\dfrac{{\rm d}}{{\rm d}t}\left(\dfrac{1}{2}\,m\,V^2\right)\,,
\end{equation}
the time integral of the power used for acceleration of the centre of mass is the change of its kinetic energy.
Therefore, to include the effect of accelerations, per lap, we need to add the increases in kinetic energy.
This is an estimate of the error committed by neglecting accelerations in expression~(\ref{eq:power}) to be quantified, for the constant-cadence and constant-power cases, in Appendix~\ref{sec:Energy}.

\begin{figure}[h]
\centering
\includegraphics[scale=0.5]{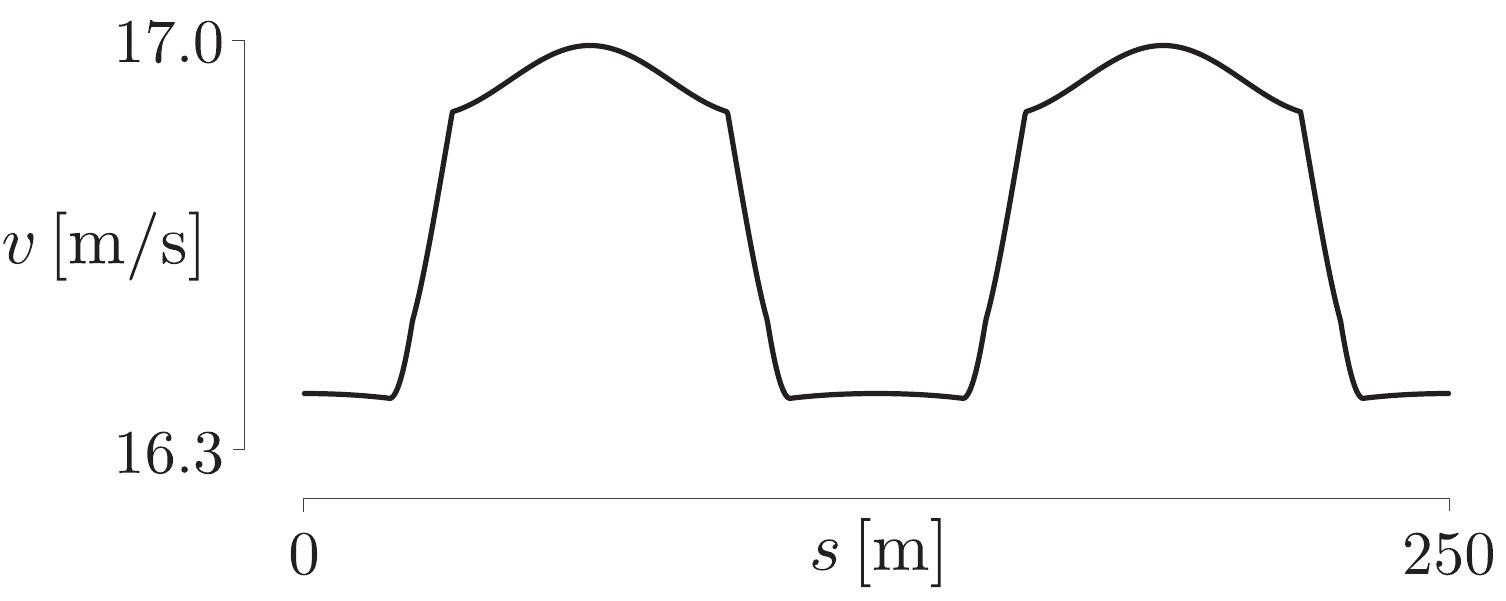}
\caption{\small Black-line speed,~$v$\,, as a function of the black-line distance,~$s$\,, for constant power}
\label{fig:FigBLspeed}
\end{figure}

The values of $v$\,, in accordance with expression~(\ref{eq:vV}), are shown in Figure~\ref{fig:FigBLspeed}, where\,---\,as expected for a constant power\,---\,leaning into the turn entails a significant increase of the black-line speed; note the difference of vertical scale between Figures~\ref{fig:FigCoMSpeed2} and \ref{fig:FigBLspeed}.
The averages are $\overline V=16.3316$\,m/s and $\overline v=16.7071$\,m/s\,.
These averages are similar to the case of the constant black-line speed averages.
Hence, maintaining a constant cadence or a constant power results in nearly the same laptime, namely, $14.9701$\,s and $14.9670$\,s\,, respectively.

To conclude this section, let us calculate the corresponding work per lap.
The work performed during a time interval, $t_2-t_1$\,, is
\begin{equation}
\label{eq:WorkConstPow}
W=\int\limits_{t_1}^{t_2}\!P\,{\rm d}t=P\!\int\limits_{t_1}^{t_2}\!{\rm d}t=P\,\underbrace{(t_2-t_1)}_{t_\circlearrowleft}=P\,t_\circlearrowleft\,,	
\end{equation}
where, for the second equality sign, we use the constancy of~$P$\,; also, we let the time interval to be a laptime.
Thus, given $\overline P=580.5941$\,W and $t_\circlearrowleft=14.9670$\,s\,, we obtain $W=8689.7680$\,J\,.

The empirical adequacy of the assumption of a constant power can be corroborated\,---\,apart from the measured power itself\,---\,by comparing experimental data to measurable quantities entailed by theoretical formulations.
The black-line speed,~$v$\,, shown in Figure~\ref{fig:FigBLspeed}, which we take to be tantamount to the wheel speed, appears to be the most reliable quantity.
Other quantities\,---\,not measurably directly, such as the centre-of-mass speed and power expended to increase potential energy\,---\,are related to $v$ by equations~(\ref{eq:PConst}) and (\ref{eq:Vvar}).

To conclude Sections~\ref{sub:ConstCad} and \ref{sub:ConstPower}, let us state that if---for the latter---we use the power obtained from the latter, we obtain the black-line speed of the former, as expected.
\section{Empirical adequacy}
\label{sec:Adequacy}
To gain an insight into empirical adequacy of the model, let us examine Section~\ref{sec:InstPower}, in the context of measurements~(Mehdi Kordi, {\it pers.~comm.}, 2020).
To do so, we use two measured quantities: cadence and force applied to the pedals, both of  which are measured by sensors attached to the bicycle.
They allow us to calculate power, which is the product of the circumferential pedal speed\,---\,obtained from cadence, given a crank length\,---\,and the force applied to pedals.
\begin{figure}[h]
\centering
\includegraphics[scale=0.35]{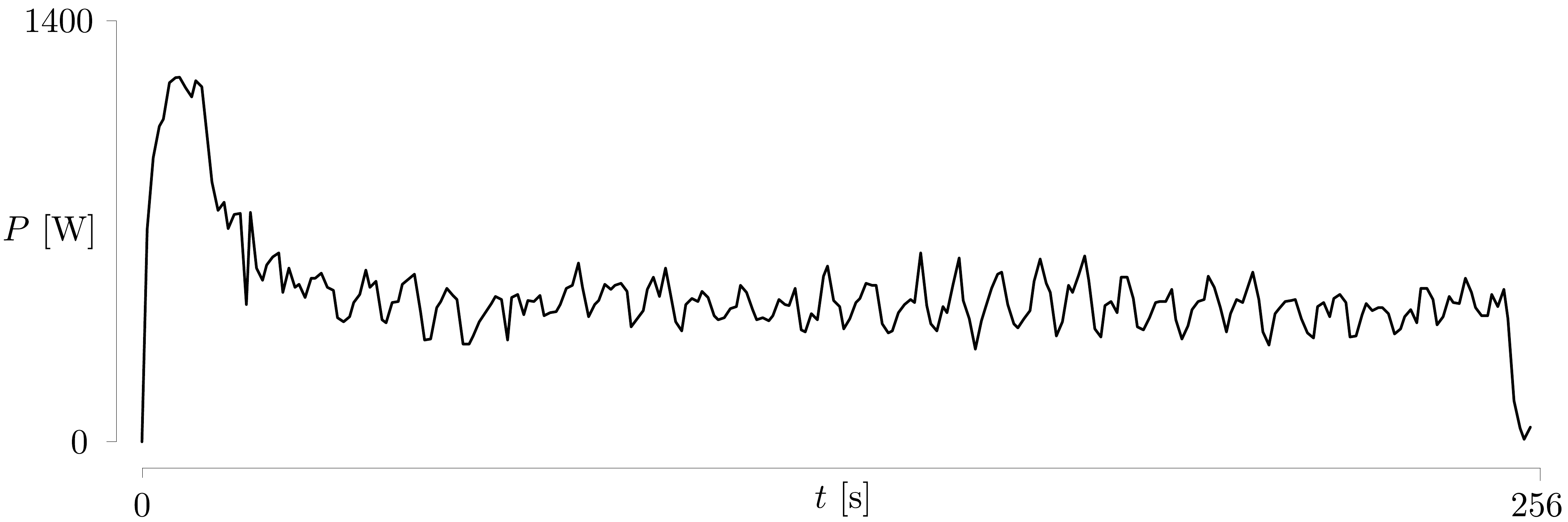}
\caption{\small Measured power,~$P$\,, as a function of the pursuit time,~$t$}
\label{fig:FigIPPower}
\end{figure}

\begin{figure}[h]
\centering
\includegraphics[scale=0.35]{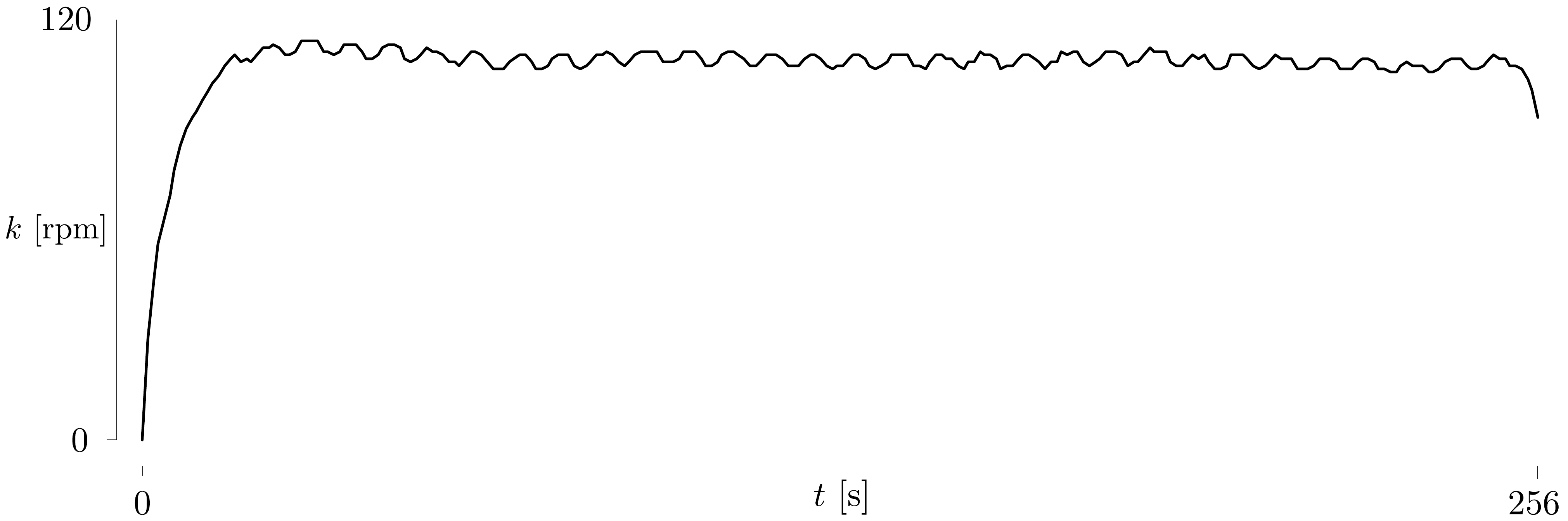}
\caption{\small Measured cadence,~$k$\,, in revolutions per minute, [rpm], as a function of the pursuit time,~$t$}
\label{fig:FigIPCadence}
\end{figure}

The measurements of power, shown in Figure~\ref{fig:FigIPPower}, oscillate about a nearly constant value, except for the initial part, which corresponds to acceleration, and the final part, where the cyclist begins to decelerate.
These oscillations are due to the repetition of straights and curves along a lap.
In particular, Figure~\ref{fig:FigIPCadence}, below, exhibits a regularity corresponding to thirty-two curves along which the cadence, and\,---\,equivalently\,---\,the wheel speed, reaches a maximum.
There are also fluctuations due to measurement errors.
A comparison of Figures~\ref{fig:FigIPPower} and \ref{fig:FigIPCadence} illustrates that power is necessarily more error sensitive than cadence, since the cadence itself is used in calculations to obtain power.
This extra sensitivity is due to intrinsic difficulties of the measurement of applied force and, herein, to the fact that values are stated at one-second intervals only, which makes them correspond to different points along the pedal rotation \citep[see also][Appendix~A]{DSSbici1}.
To diminish this effect, it is common to use a moving average, with a period of several seconds, to obtain the values of power.

To use the model to relate power and cadence, let us consider a $4000$\,-metre individual pursuit.
The model parameters are $h=1.1~\rm{m}$\,, $m=85.6~\rm{kg}$\,, ${\rm C_{d}A}=0.17~\rm{m^2}$\,, ${\rm C_{rr}}=0.0017$\,, ${\rm C_{sr}}=0.0025$\,, $\lambda=0.02$\,, $g=9.81~\rm{m/s^2}$\,, $\rho=1.17~\rm{kg/m^3}$\,.
If we use, as input, $P=488.81~\rm{W}$\,---\,which is the average of values measured over the entire pursuit\,---\,the retrodiction provided by the model results in $\overline{v}=16.86~\rm{m/s}$\,.

Let us compare this retrodiction to measurements using the fact that\,---\,for a fixed-wheel drivetrain\,---\,cadence allows us to calculate the bicycle wheel speed.
The average of the measured cadence, shown in Figure~\ref{fig:FigIPCadence}, is $k=106.56~\rm rpm$\,, which\,---\,given the gear of $9.00~\rm m$\,, over the pursuit time of $256~\rm s$\,---\,results in a distance of $4092~\rm m$\,.
Hence, the average wheel speed is~$15.98~\rm{m/s}$\,.%
\footnote{The average wheel speed is distinct from the average black-line speed,~$\overline{v}=15.63~\rm{m/s}$\,.}

The average values of the retrodicted and measured speeds appear to be sufficiently close to each other to support the empirical adequacy of our model, for the case in which its assumptions, illustrated in Figure~\ref{fig:FigBlackLine}\,---\,namely, a constant aerodynamic position and  the trajectory along the black line\,---\,are, broadly speaking, satisfied.

Specifically, they are not satisfied on the first lap, during the acceleration.
Nor can we expect them to be fully satisfied along the remainder of the pursuit, as illustrated by $4092\,{\rm m}>4000\,{\rm m}$\,, which indicates the deviation from the black-line trajectory.
\begin{figure}[h]
\centering
\includegraphics[scale=0.35]{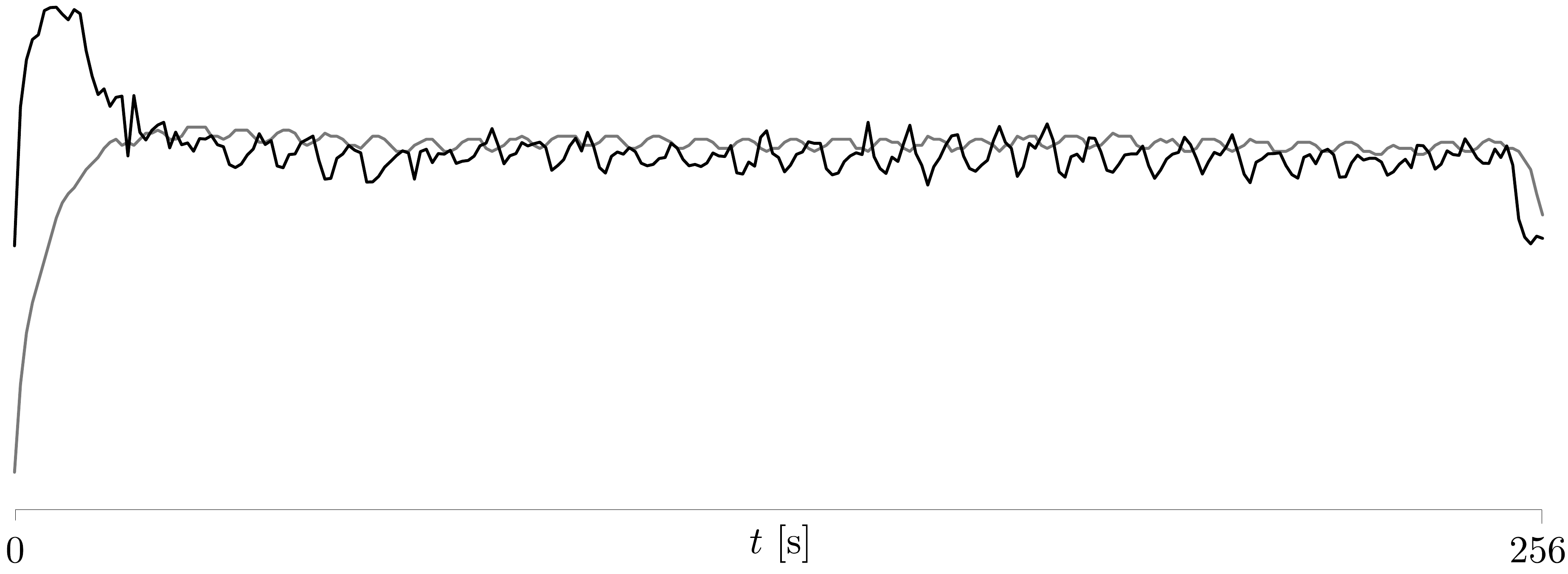}
\caption{\small Scaled values of power (black) and cadence (grey) as functions of the pursuit time,~$t$}
\label{fig:FigIPShift}
\end{figure}

\begin{figure}[h]
\centering
\includegraphics[scale=0.5]{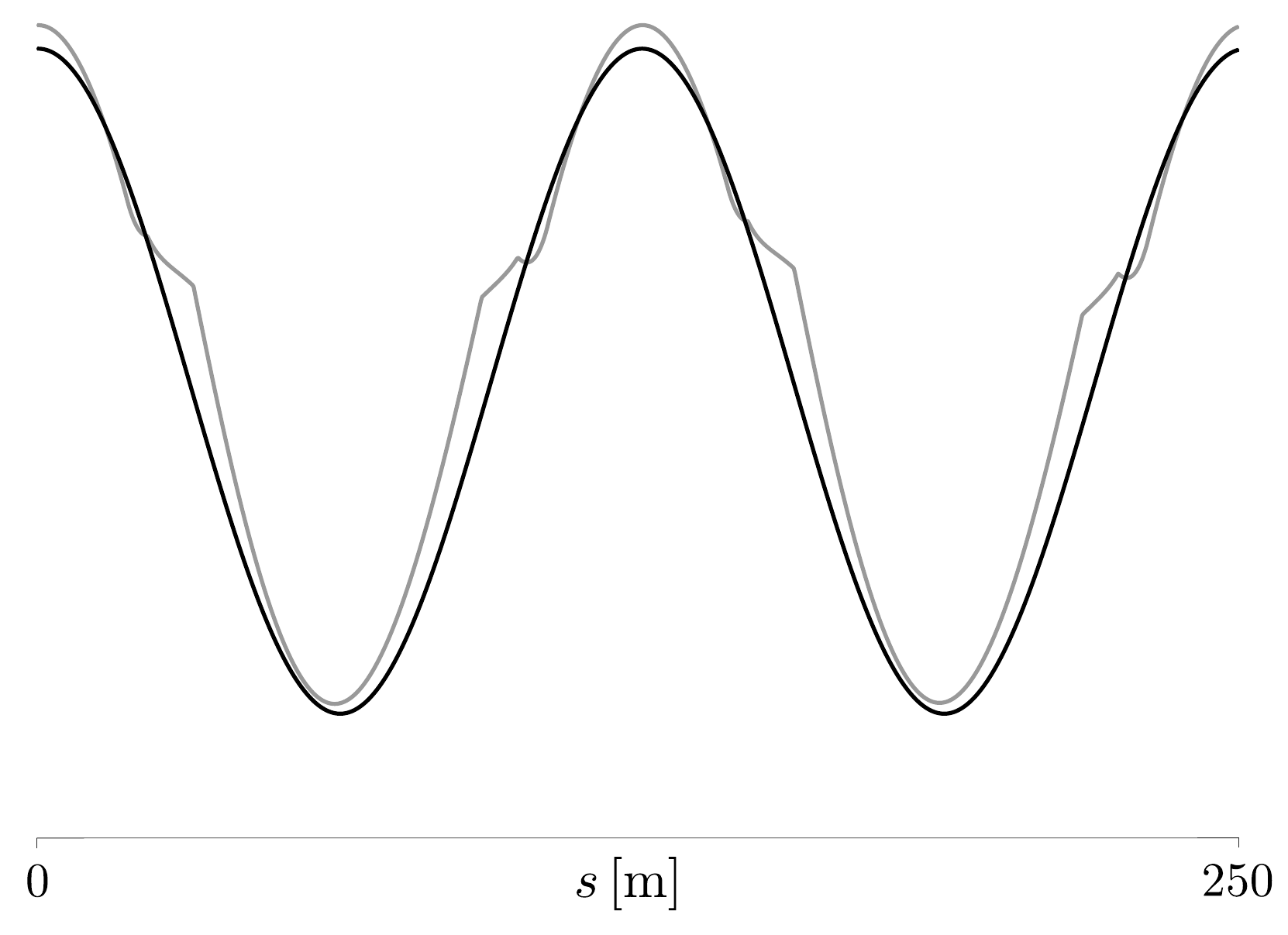}
\caption{\small Scaled values of power (black) and speed (grey) as functions of the black-line distance,~$s$}
\label{fig:FigModelShift}
\end{figure}

Furthermore, Figure~\ref{fig:FigIPShift}, which is a superposition of values scaled from Figures~\ref{fig:FigIPPower} and \ref{fig:FigIPCadence}, illustrates a shift of oscillations between power and cadence.
However, Figure~\ref{fig:FigModelShift} does not exhibit any shift.
Therein, as input, we use simulated values of power along a lap\,---\,in a manner consistent with the measured power\,---\,as opposed to single value of an average.
Thus, according to the model, the power and the black-line speed\,---\,whose pattern within the model for a fixed-wheel drivetrain is the same as for cadence\,---\,exhibit no shift.

The shift observed in Figure~\ref{fig:FigIPShift} could be an effect of measurements for a fixed-wheel drivetrain, since the value at each instant is obtained from the product of measurements of~$f_{\circlearrowright}$\,, which is the force applied to pedals, and~$v_{\circlearrowright}$\,, which is the circumferential speed of the pedals \citep[e.g.,][expression~(1)]{DSSbici1},
\begin{equation}
\label{eq:PowerMeter}
P=f_{\circlearrowright}\,v_{\circlearrowright}\,.
\end{equation}
For a fixed-wheel drivetrain, there is a one-to-one relation between $v_{\circlearrowright}$ and the wheel speed.
Hence\,---\,in contrast to a free-wheel drivetrain, for which $f_{\circlearrowright}\to0\implies v_{\circlearrowright}\to0$\,---\,the momentum of a launched bicycle-cyclist system might contribute to the value of~$v_{\circlearrowright}$\,, which is tantamount to contributing to the value of cadence.
This issue is addressed in Appendix~\ref{sec:Fixed}.

Nevertheless, the agreement between the average values of the retrodiction and measurements appears to be satisfactory.
Notably, excluding the first and last laps would increase this agreement.
For instance, if we consider, say, $33\,{\rm s} < t < 233\,{\rm s}$\,, which does not even correspond to the beginning or the end of any lap, the average power and cadence are $455.02~{\rm W}$ and $108.78~{\rm rpm}$\,, respectively.
Hence, the retrodicted and measured speeds are $16.45~\rm{m/s}$ and $16.32~\rm{m/s}$\,, respectively.
\begin{figure}[h]
\centering
\includegraphics[scale=0.35]{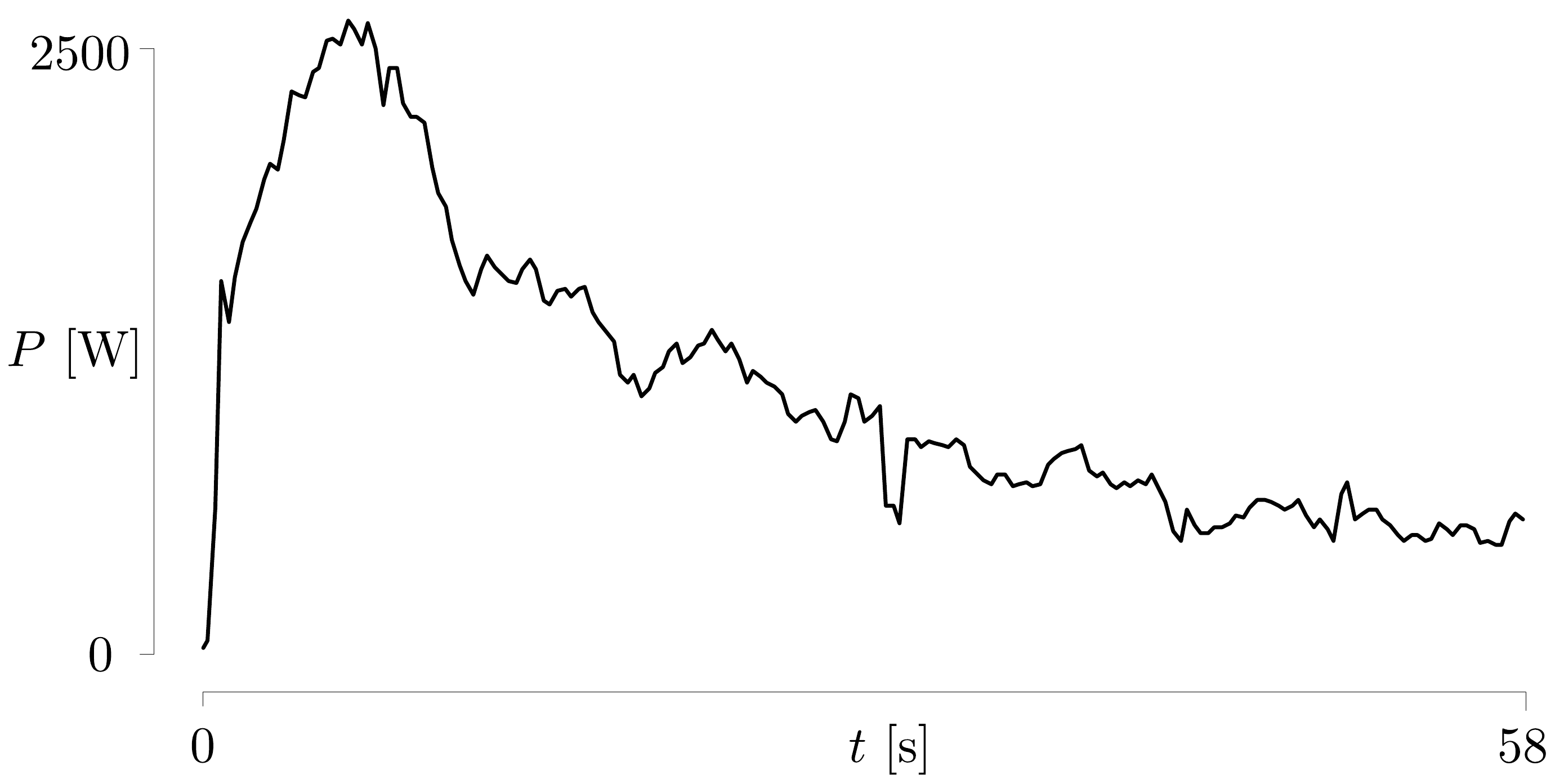}
\caption{\small Measured power,~$P$\,, as a function of the `kilo' time,~$t$}
\label{fig:FigKiloPower}
\end{figure}
\begin{figure}[h]
\centering
\includegraphics[scale=0.35]{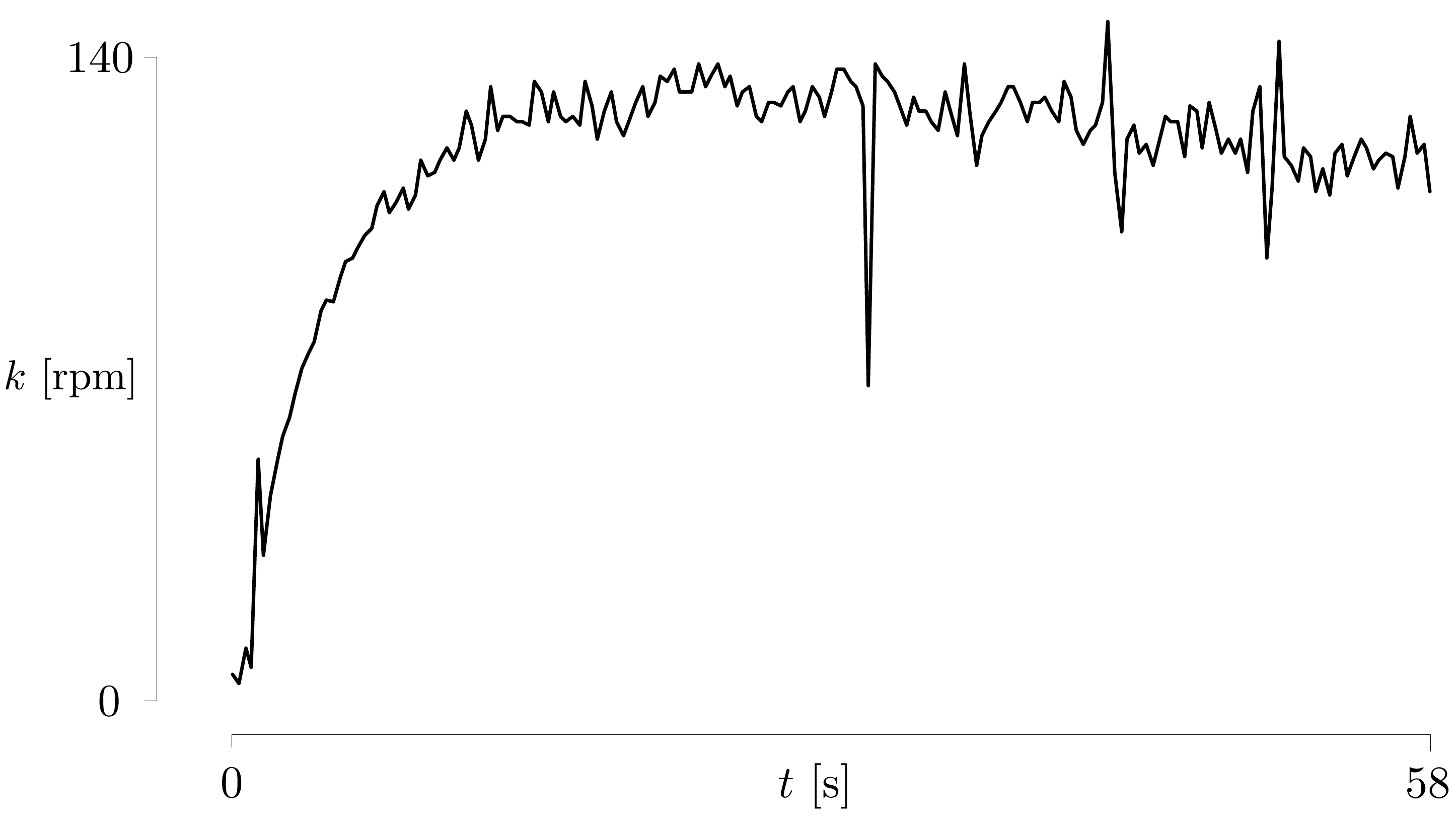}
\caption{\small Measured cadence,~$k$\,, in revolutions per minute, [rpm], as a function of the `kilo' time,~$t$}
\label{fig:FigKiloCadence}
\end{figure}

To illustrate limitations of the model, Figures~\ref{fig:FigKiloPower} and \ref{fig:FigKiloCadence} represent measurements for which it is not empirically adequate.
As shown in these figures, in this $1000$\,-metre time trial, commonly referred to as a `kilo', the cyclist reaches a steady cadence\,---\,and speed\,---\,with an initial output of power, in a manner similar to the one shown in Figure~\ref{fig:FigIPPower}.
Subsequently, in a manner similar to the one shown in Figure~\ref{fig:FigIPCadence}, the cadence remains almost unchanged, for the remainder of the time trial.
However, in contrast to Figure~\ref{fig:FigIPPower}, the power decreases.
Herein, as discussed by \citet[Appendix~B.2]{DSSbici2}, the cadence, as a function of time, is a consequence of both the power generated by a cyclist\,---\,at each instant\,---\,and the momentum of the moving bicycle-cyclist system, gained during the initial acceleration, which propels the pedals.
There is no dynamic equilibrium between the instantaneous power generated by a cyclist and the resulting cadence, in contrast to an equilibrium reached during a steady effort of a $4000$\,-metre individual pursuit.
We consider here a dynamic equilibrium {\it sensu lato}\,; the values of power and cadence, in Figures~\ref{fig:FigIPPower} and \ref{fig:FigIPCadence}, oscillate about means that are nearly constant.

In view of this equilibrium, Figures~\ref{fig:FigIPPower} and \ref{fig:FigIPCadence} would remain similar for a free-wheel drivetrain, provided a cyclists keeps on pedalling in a continuous and steady manner.
Figures~\ref{fig:FigKiloPower} and \ref{fig:FigKiloCadence} would not.
In particular, Figure~\ref{fig:FigKiloCadence} would show a decrease of cadence with time, even though the bicycle speed might not decrease significantly.
\section{Discussion and conclusions}
\label{sec:DisCon}
The mathematical model presented in this article offers the basis for a quantitative study of individual time trials on a velodrome.
The model can be used to predict or retrodict the laptimes, from the measurements of power, or to estimate the power from the recorded times.
Comparisons of such predictions or retrodictions with the measurements of time, speed, cadence and power along the track offer an insight into the empirical adequacy of a model.
Given a satisfactory adequacy and appropriate measurements, the model lends itself to estimating  the rolling-resistance, lateral-friction, air-resistance and drivetrain-resistance coefficients.
One can examine the effects of power on speed and {\it vice versa}, as well as of other parameters, say, the effects of air resistance on speed.
One can also estimate the power needed for a given rider to achieve a particular result.

In Sections~\ref{sec:InstPower}, \ref{sec:NumEx} and \ref{sec:Adequacy}, we neglect the vertical motion of the centre of mass and assume its trajectory to be contained in a horizontal plane.
In Appendix~\ref{sec:Energy}, we calculate the work done in raising and accelerating the centre of mass.
We can conclude that\,---\,even though most of the work of the cyclist is done to overcome dissipative forces\,---\,a nonnegligible portion goes into increasing mechanical energy.
This conclusion, however, does not mean that we cannot invoke expressions~(\ref{eq:LeanAngle}) and (\ref{eq:vV}), wherein we assume that the centre-of-mass trajectory is contained in a horizontal plane.
Approximations resulting from using these expressions appear to have a lesser effect on results of the model than neglecting increases of mechanical energy, which are not taken explicitly into account within model~(\ref{eq:power}). 

Presented results allow us to comment on aspects of the velodrome design.
As illustrated in Figures~\ref{fig:FigLeanAngle}--\ref{fig:FigPower}, \ref{fig:FigCoMSpeed2}, \ref{fig:FigBLspeed}, \ref{fig:FigModelShift}, the transitions\,---\,between the straights and the circular arcs\,---\,do not result in smooth functions for the lean angles, speeds and powers.
It might suggest that a commonly used Euler spiral, illustrated in Figure~\ref{fig:FigCurvature}, is not the optimal transition curve.
Perhaps, the choice of a transition curve should consider such phenomena as the jolt, which is the temporal rate of change of acceleration.
It might also suggest the necessity for the lengthening of the transition curve.

Furthermore, an optimal velodrome design would strive to minimize the separation between the zero line and the curve in Figure~\ref{fig:FigAngleDiff}, which is tantamount to optimizing the track inclination to accommodate the lean angle of a rider.
The smaller the separation, the smaller the second summand in term~(\ref{eq:modelB}).
As the separation tends to zero, so does the summand.

These considerations are to be examined in future work.
Also, the inclusion, within the model, of a change of kinetic and potential energy for instantaneous power, discussed in Appendix~\ref{sec:Energy}, is an issue to be addressed.
Another consideration to be examined is the discrepancy between the model and measurements with respect to the shift between power and cadence, illustrated in Figures~\ref{fig:FigIPShift} and \ref{fig:FigModelShift}.
A venue for such a study is introduced in Appendix~\ref{sec:Fixed}.

In conclusion, let us emphasize that our model is phenomenological.
It is consistent with\,---\,but not derived from\,---\,fundamental concepts.
Its purpose is to provide quantitative relations between the model parameters and observables.
Its justification is the agreement between measurements and predictions or retrodictions, as illustrated in Section~\ref{sec:Adequacy} by the relation between power and speed.
\section*{Acknowledgements}
We wish to acknowledge Mehdi Kordi, for information on the track geometry, used in Section~\ref{sec:Formulation}, and for measurements, used in Section~\ref{sec:Adequacy};
Tomasz Danek, for statistical insights into these measurements;
Elena Patarini, for her graphic support;
Roberto Lauciello, for his artistic contribution;
Favero Electronics for inspiring this study by their technological advances of power meters.
\section*{Conflict of Interest}
The authors declare that they have no conflict of interest.
\bibliographystyle{apa}
\bibliography{BSSSbici4.bib}

\begin{thebibliography}{}

\bibitem[\protect\astroncite{Bos et~al.}{2020}]{BSSbici6}
Bos, L., Slawinski, M.~A., and Stanoev, T. (2020).
\newblock On maximizing {VAM} for a given power: Slope, cadence, force and
  gear-ratio considerations.
\newblock {\em ar{X}iv}, 2006.15816 [physics.pop-ph].

\bibitem[\protect\astroncite{Danek et~al.}{2020a}]{DSSbici1}
Danek, T., Slawinski, M.~A., and Stanoev, T. (2020a).
\newblock On modelling bicycle power-meter measurements:~{P}art~{I}.
  {E}stimating effects of air, rolling and drivetrain resistance.
\newblock {\em ar{X}iv}, 2005.04229 [physics.pop-ph].

\bibitem[\protect\astroncite{Danek et~al.}{2020b}]{DSSbici2}
Danek, T., Slawinski, M.~A., and Stanoev, T. (2020b).
\newblock On modelling bicycle power-meter measurements:~{P}art~{II}.
  {R}elations between rates of change of model quantities.
\newblock {\em ar{X}iv}, 2005.04480 [physics.pop-ph].

\bibitem[\protect\astroncite{Slawinski et~al.}{2020}]{SSSbici3}
Slawinski, M.~A., Slawinski, R., and Stanoev, T. (2020).
\newblock On modelling bicycle power-meter measurements for velodromes: Motion
  of center of mass for individual pursuits.
\newblock {\em ar{X}iv}, 2005.04691 [physics.pop-ph].

\bibitem[\protect\astroncite{Van~Fraassen}{1980}]{Fraassen}
Van~Fraassen, B.~C. (1980).
\newblock {\em The Scientific Image}.
\newblock Oxford University Press.

\end{thebibliography}
\begin{appendix}
\section{Change of kinetic and potential energy}
\label{sec:Energy}
\setcounter{equation}{0}
\setcounter{figure}{0}
\renewcommand{\theequation}{\Alph{section}.\arabic{equation}}
\renewcommand{\thefigure}{\Alph{section}\arabic{figure}}
\subsection{Formulation}
In Sections~\ref{sec:InstPower} and \ref{sec:NumEx}, we formulate and calculate\,---\,in terms of expressions~(\ref{eq:power}) and (\ref{eq:PConst})\,---\,the power expended and work done by the cyclist against dissipative forces, namely, rolling friction, lateral friction and air resistance.
In Section~\ref{sec:Adequacy}, we examine the empirical adequacy of these expressions.
We do not consider explicitly changes in the cyclist's kinetic and potential energy, even though\,---\,assuming a constant black-line speed together with a black line contained  within a horizontal plane\,---\,the speed and height of the centre of mass change along the curved segments of the track, resulting in changes in kinetic and potential energy.

In this appendix, we calculate the work the cyclist performs to effect these changes.
In the case of a constant black-line speed, the work done to increase the system's mechanical energy can be added to the work done against the dissipative forces.
In the case of constant power, the work done to increase the system's mechanical energy cannot be so added, but only estimated approximately; strictly speaking, including that work contradicts the assumption of constant power.
In general, to include the work done to increase mechanical energy would require modifying the model for instantaneous power, stated in expression~(\ref{eq:power}).

Consider a cyclist following a track whose curvature is $R_1$\,, with a centre-of-mass speed,~$V_1$\,.
We wish to determine the work the cyclist must perform to change the kinetic and potential energy when the radius of curvature and speed change to new values,~$R_2$ and $V_2$\,, respectively.
Let us first determine the work required to change the kinetic energy.

Neglecting the kinetic energy of the rotating wheels, the kinetic energy of the system is due to the translational motion of the centre of mass,
\begin{equation}
\label{eq:KinEn}
    K = \tfrac{1}{2}\,m\,V^2\,.
\end{equation}
To find the work done to change kinetic energy, it is important to note that the bicycle-cyclist system is not purely mechanical, in the sense that in addition to kinetic and potential energy the system also possesses internal energy in the form of the cyclist's chemical energy.
When the cyclist speeds up, internal energy is converted into kinetic.
However, the converse is not true: when the cyclist slows down, kinetic energy is not converted into internal.
It follows that, to determine the work the cyclist does to change kinetic energy, we should consider only the increases.
Then, if the centre-of-mass speed increases monotonically from $V_1$ to $V_2$\,, the work done is the increase in kinetic energy,
\begin{equation*}
    \Delta K = \tfrac{1}{2}\,m\left(V_2^2-V_1^2\right)\,.
\end{equation*}
The stipulation of a monotonic increase is needed due to the nonmechanical nature of the system, with the cyclist doing positive work\,---\,with an associated decrease in internal energy\,---\,when speeding up, but not doing negative work\,---\,with an associated increase in internal energy\,---\,when slowing down.
Put another way, the cyclist's work to increase speed from $V_1$ to $V_2$ depends not only on the initial and final speeds but also on the intermediate speeds.

Let us determine the work required to change the potential energy,
\begin{equation}
\label{eq:PotEn}
    U = m\,g\,h\,\cos\vartheta\,.
\end{equation}
The lean angle\,, $\vartheta$\,, is determined by assuming that, at all times, the system is in rotational equilibrium about the line of contact of the tires with the ground; this assumption yields the implicit condition on $\vartheta$\,, stated in expression~(\ref{eq:LeanAngle}).
Since the lean angle depends on the centre-of-mass speed,~$V$\,, as well as the radius of curvature of the track,~$R$\,, it\,---\,and therefore the height of the centre of mass\,---\,change if either $V$ or $R$ changes.
The work done is the increase in potential energy resulting from a monotonic decrease in the lean angle from $\vartheta_1$ to $\vartheta_2$\,,
\begin{equation*}
    \Delta U = m\,g\,h\,\left(\cos\vartheta_2-\cos\vartheta_1\right)\,,
\end{equation*}
as can be seen in Figure~\ref{fig:FigLeanAngle}.
The same considerations\,---\,due to the nonmechanical nature of the system\,---\,apply to the work done to change potential energy as to that done to change kinetic energy.
The cyclist does positive work\,---\,with an associated decrease in internal energy\,---\,when straightening up, but not negative work\,---\,with an associated increase in internal energy\,---\,when leaning into the turn.
And as before, the cyclist's work to decrease in the lean angle from $\vartheta_1$ to $\vartheta_2$ depends not only on the initial and final lean angles but also on the intermediate angles.

The changes in potential energy due to a changing lean angle are different in character from the changes associated with hill climbs, stated as the first term in the numerator of expression~(1) of \citet{SSSbici3}.
When climbing a hill, a cyclist does work to increase potential energy.
When descending, at least some of that potential energy is converted into kinetic energy of forward motion.
This is not the case with the work the cyclist does to straighten up.
When the cyclist leans, potential energy is not converted into kinetic energy of forward motion.
\subsection{Constant cadence}
\label{sub:ConCad}
\begin{figure}[h]
\centering
\includegraphics[scale=0.5]{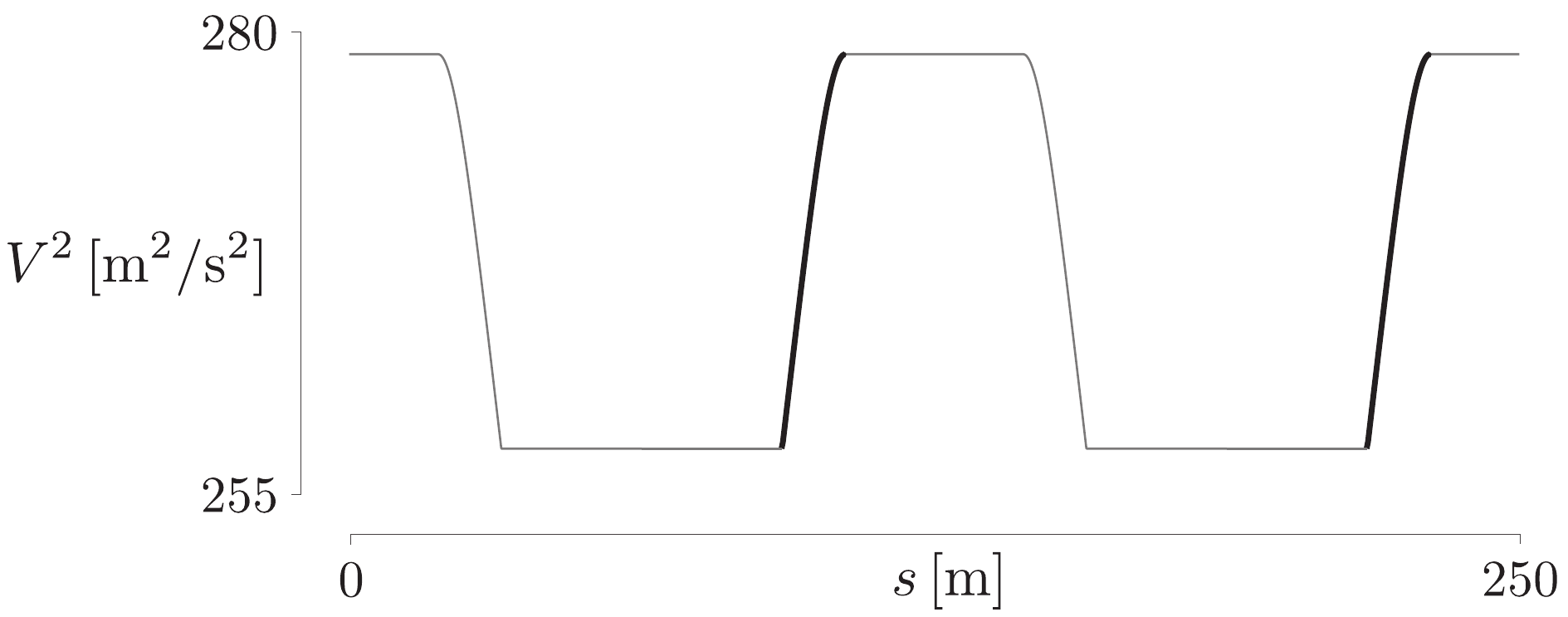}
\caption{\small Increases of~$V^2$\,, as a function of the black-line distance,~$s$\,, for constant cadence}
\label{fig:FigIncVSqCad}
\end{figure}

In the example considered in Section~\ref{sub:ConstCad}, assuming a constant cadence, which is equivalent to a constant black-line speed, the sum of increases of the centre-of-mass speed squared over one lap, shown by thick black lines in Figure~\ref{fig:FigIncVSqCad}, is $\sum\Delta V^2=42.6148\, {\rm m^2/s^2}$\,.
This results in an increase in kinetic energy of
\begin{equation*}
    \Delta K = \tfrac{1}{2}\,m\sum\Delta V^2 = 1789.8216\,{\rm J}\,,
\end{equation*}
per lap.
Invoking the equivalence of the energy increase and work\,---\,in view of the laptime, $t_\circlearrowleft=14.9701\,{\rm s}$\,, and in accordance with expression~(\ref{eq:WorkConstCad})\,, including the drivetrain-resistance coefficient, $\lambda=0.02$\,---\,the average power required for this increase is
\begin{equation}
\label{eq:PK}
P_K=\dfrac{1}{1-\lambda}\dfrac{\Delta K}{t_\circlearrowleft}=122.0000\,{\rm W}\,.
\end{equation}
In the same example, the sum of increases of the centre-of-mass height over one lap is $\sum h\,\Delta\cos\vartheta=0.8390\,{\rm m}$\,, which results in an increase in potential energy of
\begin{equation*}
    \Delta U = m\,g\,h\sum\Delta\cos\vartheta = 691.3486\,{\rm J}
\end{equation*}
and
\begin{equation}
\label{eq:PU}
P_U=\dfrac{\Delta U}{t_\circlearrowleft}=46.1820\,{\rm W}\,;
\end{equation}
the power resulting in the increase of potential energy is not affected by the drivetrain resistance.
Adding $P_K$ and $P_U$\,, we find $168.1820\,{\rm W}$\,, which is the average power, per lap, required to increase the mechanical energy.
\subsection{Constant power}
\label{sub:ConPow}
\begin{figure}[h]
\centering
\includegraphics[scale=0.5]{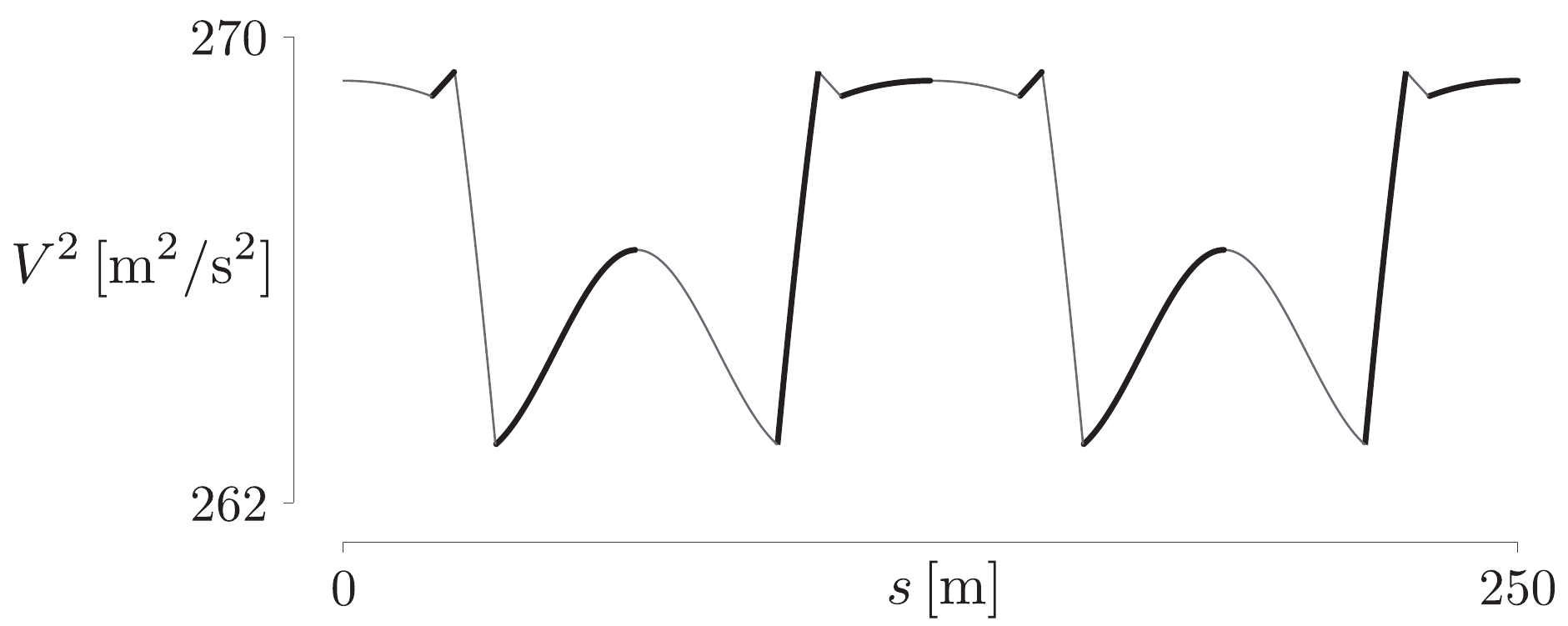}
\caption{\small Increases of~$V^2$\,, as a function of the black-line distance,~$s$\,, for constant power}
\label{fig:FigIncVSqPower}
\end{figure}
In the example considered in Section~\ref{sub:ConstPower}, we assume a constant instantaneous power needed to overcome dissipative forces, without including explicitly changes in mechanical energy.
The sum of increases of the centre-of-mass speed squared over one lap, shown by thick black lines in Figure~\ref{fig:FigIncVSqPower}, is $\sum\Delta V^2=20.9662\,{\rm m^2/s^2}$\,, which results in an increase in kinetic energy of
\begin{equation*}
    \Delta K = \tfrac{1}{2}\,m\sum\Delta V^2 = 880.5804\,{\rm J}\,.
\end{equation*}
In view of the laptime, $t_\circlearrowleft=14.9670\,{\rm s}$\,, and in accordance with expression~(\ref{eq:WorkConstPow}), including the drivetrain-resistance coefficient, $\lambda=0.02$\,, the average power required for this increase is
\begin{equation*}
P_K=\dfrac{1}{1-\lambda}\dfrac{\Delta K}{t_\circlearrowleft}=60.0355\,{\rm W}\,.
\end{equation*}

In the same example, the sum of increases of the centre-of-mass height over one lap is $\sum h\Delta\cos\vartheta=0.8698\,{\rm m}$\,, which results in an increase in potential energy of
\begin{equation*}
    \Delta U = m\,g\,h\sum\Delta\cos\vartheta = 716.7249\,{\rm J}
\end{equation*}
and
\begin{equation*}
P_U=\dfrac{\Delta U}{t_\circlearrowleft}=47.8870\,{\rm W}\,.
\end{equation*}
Adding $P_K$ and $P_U$\,, we find that $107.9225\,{\rm W}$\,, which is the average power, per lap, required to increase the mechanical energy.
\subsection{Instantaneous power revisited}
Let us consider a general expression for the power required to increase the mechanical energy,
\begin{equation}
\label{eq:PowerEnergy}
P=
\underbrace{\,m\,g\,h\left(\dfrac{{\rm d}\cos\vartheta}{{\rm d}t}\right)_+}_{P_U}+
\underbrace{\dfrac{V}{1-\lambda}\,m\left(\dfrac{{\rm d}V}{{\rm d}t}\right)_+}_{P_K}	
\end{equation}
where $(\,\cdot\,)_+:=\max(\,\cdot\,,0)$ denotes the positive part of the quantity in parentheses,
and $t$ stands for time; the relation between $V$ and $\vartheta$ is stated in expression~(\ref{eq:Vvar}).
As discussed, at any instant, only positive summands are included.

In view of expression~(\ref{eq:PotEn}),  the temporal rate of change of potential energy is $P_U:={\rm d}U/{\rm d}t=m\,g\,h\,{\rm d}\cos\vartheta/{\rm d}t$\,, which has units of the product of force and speed, as required.
The speed, however, is neither the centre-of-mass speed,~$V$\,, nor the black-line speed,~$v$\,; it is the speed with which the centre of mass rises, ${\rm d}(h\,\cos\vartheta)/{\rm d}t$\,.
The power expended by the cyclist to increase potential energy is denoted by $P_U$\,, in expression~(\ref{eq:PowerEnergy}).

In contrast to the power expended to increase the kinetic energy\,---\,which is generated by pedalling and, hence, involves the drivetrain\,---\,herein, as the cyclist slows down and the gravitational torque tends to increase the lean angle, the cyclist does work against this torque to straighten up.
The cyclist needs to perform this work even without applying any force to the pedals; hence, the drivetrain is not involved in increasing potential energy.

In view of expressions~(\ref{eq:dK}) and (\ref{eq:KinEn}), the temporal rate of change of kinetic energy is ${\rm d}K/{\rm d}t=m\,V\,{\rm d}V/{\rm d}t$\,.
Due to the drivetrain resistance, the power,~$P_K$\,, expended by the cyclist to increase kinetic energy is the solution of $m\,V\,{\rm d}V/{\rm d}t = (1 - \lambda)\,P_K$\,, which is the second term of the first line in expression~(\ref{eq:PowerEnergy}).

\begin{figure}[h]
\centering
\includegraphics[scale=0.5]{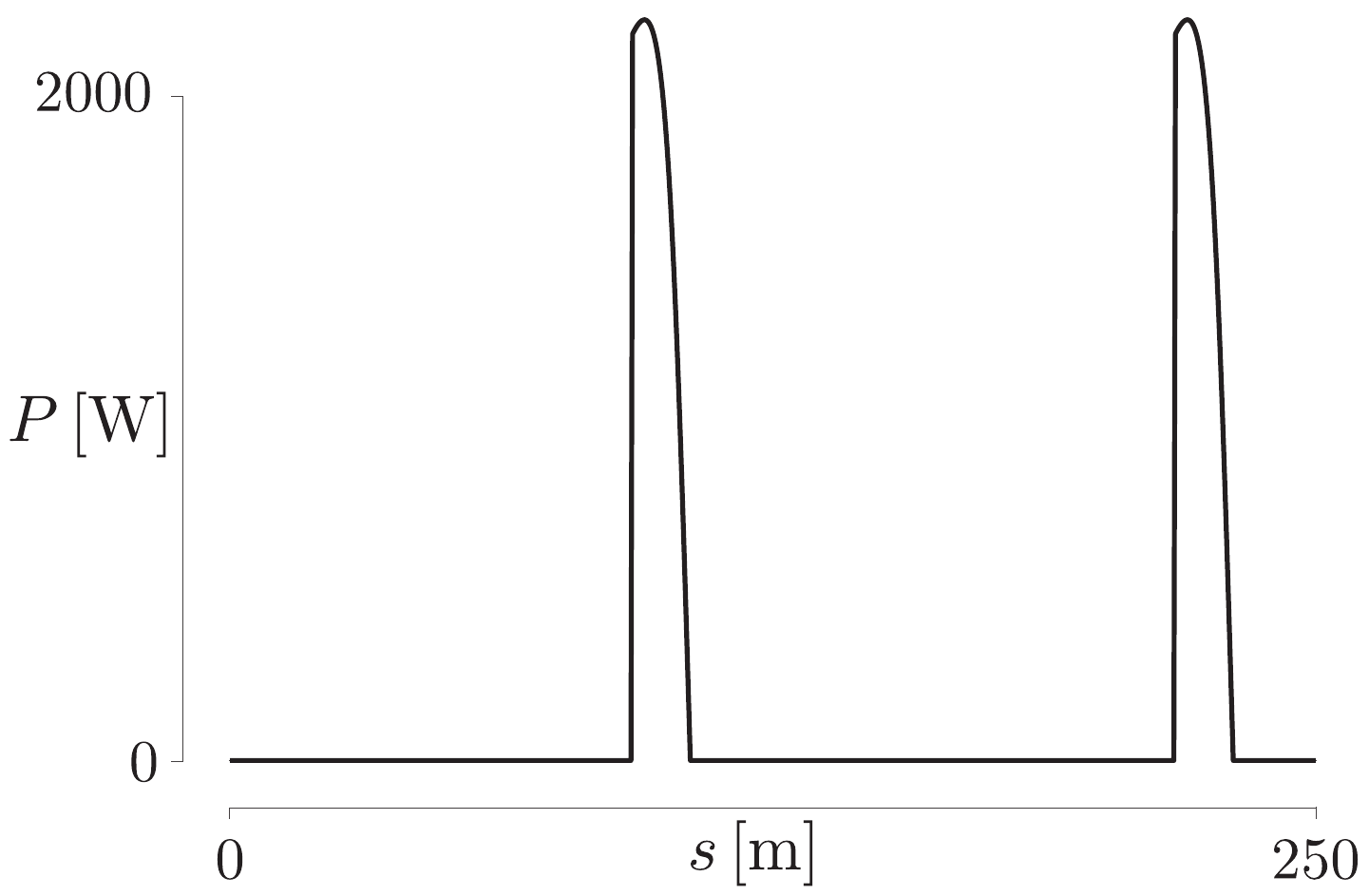}
\caption{\small{Instantaneous power to increase the kinetic and potential energy, as a function of the black-line distance,~$s$\,, for constant cadence}}
\label{fig:FigMechEn}
\end{figure}

Following expression~(\ref{eq:PowerEnergy})\,---\,using model parameters stated in Section~\ref{sub:ModPar} and a constant black-line speed,~$v=16.7\,\rm m/s$\,---\,we obtain, using $100\,000$ discrete points, the average power required for the increase of mechanical energy of $168.1905\,\rm W$\,, which, as expected, is equivalent to the sum of values stated in expressions~(\ref{eq:PK}) and (\ref{eq:PU}), namely, $168.1820\,\rm W$\,; they agree to the numerical accuracy employed.
Since this power corresponds to the rising of the centre of mass upon exiting the curve, it is confined to two transition curves, as shown in Figure~\ref{fig:FigMechEn}.

The distribution of power shown in Figure~\ref{fig:FigMechEn} is not consistent with measurements.
This inconsistency might\,---\,in part\,---\,be a consequence of our assumption of the rotational equilibrium, at all times, resulting in the fact that\,---\,in accordance with our model, and as shown in Figure~\ref{fig:FigLeanAngle}\,---\,the lean angle changes only along the transitions curves, which results in the need of power used to increase mechanical energy being restricted to these curves.

The most important, however, is the fact that the power measurements with a fixed-wheel drivetrain are affected by the momentum of the bicycle-cyclist system, as discussed by \citet[Appendix~B.2]{DSSbici2}.
Consequently, in contrast to instantaneous power shown in Figure~\ref{fig:FigMechEn}, we might consider the average values formulated in Appendices~\ref{sub:ConCad} and \ref{sub:ConPow}.
The agreement between the model and measurements, discussed in Section~\ref{sec:Adequacy}, indicates that these values are included in averages\,---\,over a lap or its multiples\,---\,of the measured power.
Hence, we infer that the power used to increase the kinetic and potential energy, stated in expressions~(\ref{eq:KinEn}) and (\ref{eq:PotEn}), respectively, is implicitly accounted for by the model stated in expression~(\ref{eq:power}).
Specifically, the power required to increase the mechanical energy is incorporated in the power to account for the drivetrain, tire and air frictions stated in expressions~(\ref{eq:modelO}), (\ref{eq:modelB}) and (\ref{eq:modelC}), respectively.
This necessarily leads to an overestimate of the values of $\lambda$\,, $\rm C_{rr}$\,, $\rm C_{sr}$ and $\rm C_{d}A$ to achieve the agreement between the model and measurements.
\section{Instantaneous measurements}
\label{sec:Fixed}
\setcounter{equation}{0}
\setcounter{figure}{0}
\setcounter{remark}{0}
Let us return to expression~(\ref{eq:power}) and consider it explicitly in the context of instantaneous measurements.
Invoking expression~(\ref{eq:PowerMeter}), we express power as the product of $f_\circlearrowright$ and $v_\circlearrowright$\,. 
In view of the fixed-wheel drivetrain, we relate $v$\,, which appears in expression~(\ref{eq:modelB}), to~$v_{\circlearrowright}$ by $v=\xi\,v_{\circlearrowright}$\,, where $\xi$ is a constant.
Thus, formally we write
\begin{align*}
\overbrace{f_{\circlearrowright}\,v_{\circlearrowright}}^P=&\\
&\dfrac{1}{1-\lambda}\,\,\Bigg\{\\
&\left.\left.\Bigg({\rm C_{rr}}\,m\,g\,(\sin\theta\tan\vartheta+\cos\theta)\cos\theta
+{\rm C_{sr}}\Bigg|\,m\,g\,\frac{\sin(\theta-\vartheta)}{\cos\vartheta}\Bigg|\sin\theta\Bigg)\,\overbrace{\xi\,v_{\circlearrowright}}^v
\right.\right.\\
&+\,\,\tfrac{1}{2}\,{\rm C_{d}A}\,\rho\,V^3\Bigg\}\,,
\end{align*}
where we assume the black-line speed and the wheel speed to be the same.
Since the measurements consist of $f_{\circlearrowright}$ and cadence,~$k$\,, we write
\begin{equation*}
v_{\circlearrowright}=2\pi\,\ell\,k\,,
\end{equation*}
where $\ell$ is the crank length.
Hence,
\begin{equation*}
\xi=\dfrac{v}{v_{\circlearrowright}}
=\dfrac{2\pi\,L\,G\,k}{2\pi\,\ell\,k}
=\dfrac{L\,G}{\ell}\,,
\end{equation*}
where $L$ and $G$ are the wheel radius and the gear ratio, respectively.
Hence,
\begin{equation}
\label{eq:vkLG}
v=\xi\,v_{\circlearrowright}=	\dfrac{L\,G}{\ell}\,2\pi\,\ell\,k
=2\pi\,k\,L\,G\,.
\end{equation}
Explicitly,
\begin{align}
\label{eq:Measurements}
2\pi\,\ell\,f_{\circlearrowright}\,k=&\\
\nonumber&\dfrac{1}{1-\lambda}\,\,\Bigg\{\\
\nonumber&\left.\left.2\pi\,L\,G\,\Bigg({\rm C_{rr}}\,m\,g\,(\sin\theta\tan\vartheta+\cos\theta)\cos\theta
+{\rm C_{sr}}\Bigg|\,m\,g\,\frac{\sin(\theta-\vartheta)}{\cos\vartheta}
\Bigg|\sin\theta\Bigg)\,k
\right.\right.\\
\nonumber&+\,\,\tfrac{1}{2}\,{\rm C_{d}A}\,\rho\,V^3\Bigg\}\,,
\end{align}
where force,~$f_{\circlearrowright}$\,, and cadence,~$k$\,, are the two measured quantities.
Other information\,---\,gravitational acceleration,~$g$\,, air density,~$\rho$\,, mass,~$m$\,, wheel radius,~$L$\,, crank length,~$\ell$\,, gear ratio,~$G$\,, resistance coefficients,~$\lambda$\,, $\rm C_{rr}$\,, $\rm C_{sr}$\,, $\rm C_dA$\,---\,need to be provided, and\,---\,given centre-of-mass height,~$h$\,, and curve radius,~$R$\,---\,$\theta$\,, $\vartheta$ and $V$ are calculated.

\begin{figure}[h]
\centering
\includegraphics[scale=0.35]{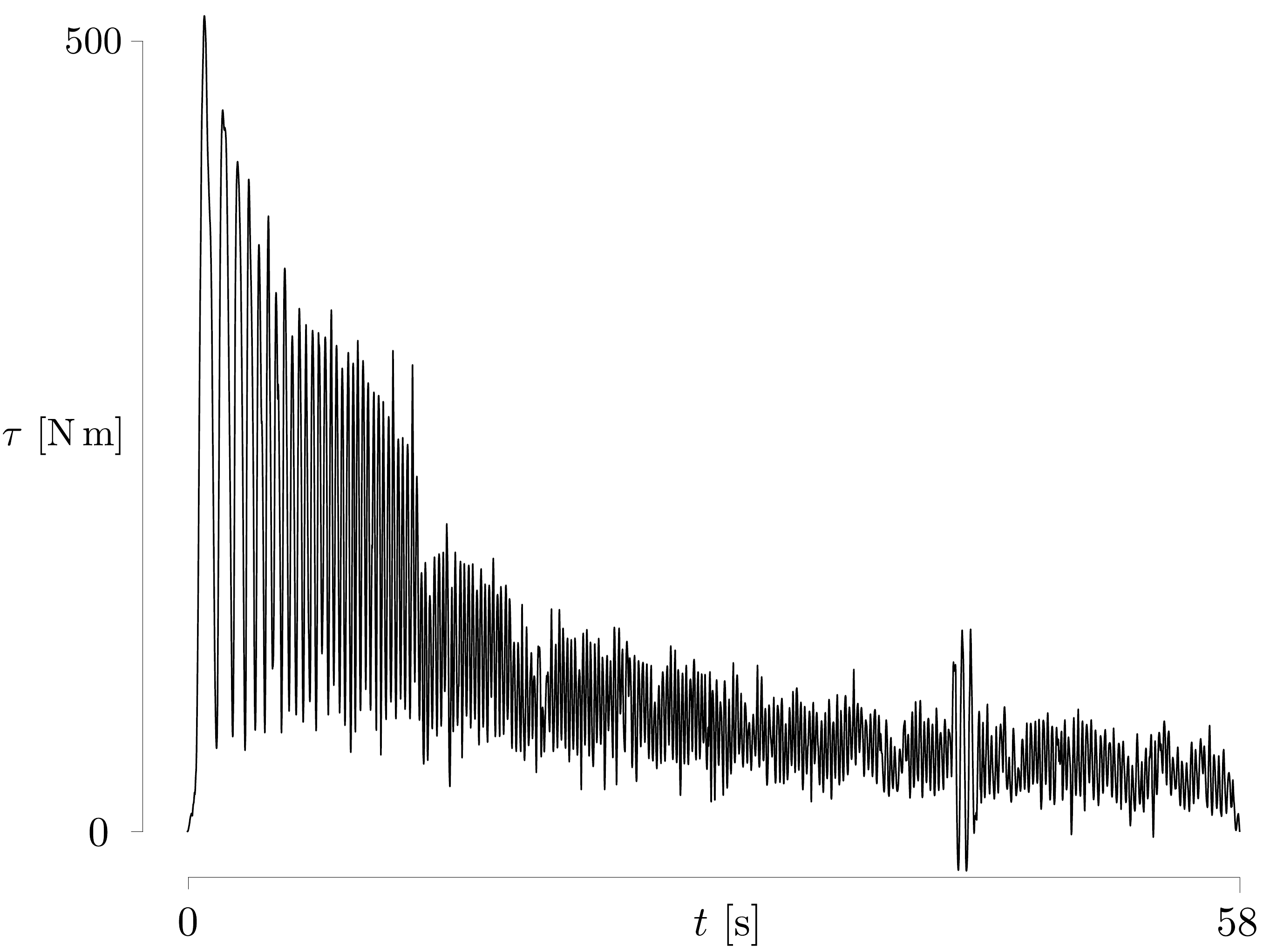}
\caption{\small{Measured torque,~$\tau$\,, as a function of the `kilo' time,~$t$}}
\label{fig:FigKiloTorque}
\end{figure}
\begin{remark}
One can write the left-hand side of expression~(\ref{eq:Measurements}) as $P=2\pi\,\tau\,k$\,, where $\tau:=\ell\,f_{\circlearrowright}$ is the torque.
The torque\,---\,shown in Figure~\ref{fig:FigKiloTorque}\,---\,corresponds to the cadence shown in Figure~\ref{fig:FigKiloCadence}\,;  herein, the torque\,---\,in contrast to cadence\,---\,is measured using a high-frequency device.
These are the two measurements whose product is the power shown in Figure~\ref{fig:FigKiloPower}\,.%
\footnote{Notably, in view of a rather steady cadence, the decrease of power is consistent with the decrease of torque, which is tantamount to the decrease of force.}
\end{remark}

Expression~(\ref{eq:Measurements}) differs from expression~(\ref{eq:power}) in several aspects, since it takes into account the method of power-meter measurements.
We recognize that power is not a direct measurement, but the product of the measured force\,---\,applied by a cyclist\,---\,and measured cadence, at that instant.
Hence\,---\,in contrast to a free-wheel drivetrain\,---\,for a fixed-wheel drivetrain, this product is not equivalent to the instantaneous power generated by a cyclist.
At a given instant, cadence is not solely a consequence of the force applied by a cyclist, at that instant; it is also a consequence of the momentum of a launched bicycle.

Furthermore, in contrast to expression~(\ref{eq:power}), in expression~(\ref{eq:Measurements}), $v$ is not obtained from the model.
 It results\,---\,due to a fixed-wheel drivetrain\,---\,directly from the measurements of cadence, given the wheel radius and gear ratio, as stated in expression~(\ref{eq:vkLG}).

Since, for a fixed-wheel drivetrain, the measured power corresponds to the power of the bicycle-cyclist system, not only to the instantaneous power generated by a cyclist, a future examination of the force and cadence measurements\,---\,based on expression~(\ref{eq:Measurements})\,---\,might allow us to gain an insight into the contribution of the momentum of the system to the measurement of cadence.
However, even without such an examination\,---\,given independent measurements of force and cadence\,---\,this expression can be used to estimate, by minimizing a misfit, the instantaneous values of model parameters, such as the air-resistance coefficient,~$\rm C_dA$\,.
One might argue that such an estimate is preferable, since $\rm C_dA$ depends directly on the power of the system, not on the instantaneous power generated by a cyclist.
Also, as discussed in Section~\ref{sec:Adequacy}\,---\,without such an examination\,---\,only the product of measurements of $f_\circlearrowright$ and $v_\circlearrowright$\,, which results in power, allows us to estimate the average, as opposed to instantaneous, values.
Such estimates result in a satisfactory agreement between the model and measurements, as exhibited in that section by the relation between an average power and an average speed, per lap.
\end{appendix}
\end{document}